\documentclass{aa}  

\usepackage{graphicx}
\usepackage{lscape}
\usepackage{rotating}
\usepackage[labelfont=bf]{caption}
\usepackage{subfig}
\usepackage{amsmath}
\usepackage{natbib}
\usepackage{txfonts}
\usepackage{hyperref}
\hypersetup{colorlinks=true,citecolor=blue,linkcolor=blue}
\newcommand{\change}[1]{#1}
\newcommand{\newchange}[1]{#1}
\newcommand{\refchange}[1]{#1}
\newcommand{\newrefchange}[1]{#1}
\pdfoutput=1

\begin{document}

   \title{Probing midplane CO abundance and gas temperature with DCO$^+$ in the protoplanetary disk around HD 169142}

%    \subtitle{....}

   \author{M.T. Carney \inst{1}, D. Fedele \inst{2}, M.R. Hogerheijde \inst{1,3}, C. Favre \inst{2},
           C. Walsh \inst{4}, S. Bruderer \inst{5}, A. Miotello \inst{6}, \\
           N.M. Murillo \inst{1}, P.D. Klaassen \inst{7},
           Th. Henning \inst{8}, E.F. van Dishoeck \inst{1,5}
          }

   \institute{Leiden Observatory, Leiden University, PO Box 9513, 2300 RA, The Netherlands \\
              \email{masoncarney@strw.leidenuniv.nl}
         \and
             INAF–Osservatorio Astrofisico di Arcetri, L.go E. Fermi 5, 50125 Firenze, Italy
         \and
	     Anton Pannekoek Institute for Astronomy, University of Amsterdam, Science Park 904, 1098 XH, Amsterdam, The Netherlands
         \and 
             School of Physics and Astronomy, University of Leeds, Leeds LS2 9JT, UK
         \and 
             Max-Planck-Institut für Extraterrestrische Physik, Giessenbachstrasse 1, 85748 Garching, Germany
         \and
	     European Southern Observatory, Garching bei M\"{u}nchen, Germany
         \and
             UK Astronomy Technology Centre, Royal Observatory Edinburgh, Blackford Hill, Edinburgh EH9 3HJ, UK
         \and 
             Max Planck Institute for Astronomy, Koenigstuhl 17, 69117 Heidelberg, Germany
             }

%    \date{in prep...}
   \date{Received November 11, 2017; accepted February 23, 2018}

% \abstract{}{}{}{}{} 
% 5 {} token are mandatory
 
  \abstract
   {
   Physical and chemical processes in protoplanetary disks affect the disk structure and the midplane
   environment within which planets form. The simple \newchange{deuterated molecular cation} DCO$^+$ 
   \newchange{has been proposed to act} as a tracer of the disk midplane conditions.
   }
   {
   This work aims to understand which midplane conditions \newchange{are probed by the DCO$^+$ emission}
   in the disk around the Herbig Ae star HD 169142. We explore the sensitivity of the DCO$^+$ 
   formation pathways to the gas temperature and the CO abundance.
   }
   {
   The DCO$^+$ $J = 3-2$ transition was observed with ALMA at a spatial resolution of $\sim$0.3$\arcsec$
   (35 AU \newchange{at 117 pc}). We modeled DCO$^+$ emission in HD 169142 with a \newchange{physical disk structure} adapted from the 
   literature, and employed a simple deuterium chemical network to investigate the formation of DCO$^+$ 
   through the cold deuterium fractionation pathway via H$_2$D$^+$. 
   Parameterized models are used to \newchange{modify} the gas temperature and 
   CO abundance structure of the \newchange{disk midplane to test} their effect on DCO$^+$ production. Contributions from the warm 
   deuterium fractionation pathway via CH$_2$D$^+$ are \newchange{approximated} using a constant 
   abundance \newchange{in the intermediate disk layers}.
   }
   {
   The DCO$^+$ line is detected in the HD 169142 disk with a total \newchange{integrated} line flux of 730$\pm$73 mJy km s$^{-1}$.
   \newchange{The radial intensity profile reveals a warm, inner component of the DCO$^+$ emission at radii $\lesssim$30 AU 
   and a broad, ring-like structure from $\sim$50 -- 230 AU with a peak at 100 AU just beyond the edge of the millimeter grain distribution.}
   Parameterized models show that alterations to the midplane gas temperature and CO abundance 
   are both needed to recover the observed DCO$^+$ radial intensity profile. \refchange{The alterations
   are relative to the fiducial physical structure of the literature model constrained by dust and CO observations.}
   \newchange{The best-fit model contains a shadowed, cold midplane in the region $z/r$ < 0.1 
   with an 8 K decrease in $T_{\rm gas}$ and a factor of five CO depletion just
   beyond the millimeter grains ($r=83$ AU), and a 2 K decrease in $T_{\rm gas}$ for $r>120$ AU.} The warm deuterium fractionation 
   pathway is implemented as a constant DCO$^+$ abundance of $2.0 \times 10^{-12}$ between 30--70 K and 
   contributes $>$85\% to the DCO$^+$ emission at $r < 83$ AU in the best-fit model.
   }
   {
   The DCO$^+$ emission probes a reservoir of cold material in the HD 169142 outer disk 
   that is not probed by the millimeter continuum, the SED, nor the emission from the $^{12}$CO, 
   $^{13}$CO, or C$^{18}$O \newchange{$J=2-1$ lines}. The DCO$^+$ emission is \newchange{a sensitive probe of gas
   temperature and CO abundance} near the disk midplane \newchange{and provides information about the outer 
   disk beyond the millimeter continuum distribution that is largely absent in \refchange{abundant gaseous tracers
   such as CO isotopologues}.}
   }

   \keywords{astrochemistry -- protoplanetary disks -- submillimeter:planetary systems}
   \authorrunning{Carney, M.~T. et al.}
   \titlerunning{DCO$^+$ in HD 169142}
   
   \maketitle
%
%________________________________________________________________

\section{Introduction}
\label{sec:intro}

Protoplanetary disks have complex structures due to the many physical
and chemical processes that shape their environment. This includes
but is not limited to radiative heating from the \newchange{central
pre-main sequence (PMS) star}, viscous heating, molecular line 
cooling, photodissociation and ionization, dust grain growth and 
radial drift, and the freeze-out of molecular species in cold disk regions
\citep[for a review of disk structure and evolution, see][]{Williams2011}.
These processes culminate in the formation of terrestrial and giant
planets, often before the gas disk is dispersed. The study of protoplanetary
disk structure while the disks still retain their large gas reservoirs is 
important to understand the environments in which planets will form.

Several gas-rich protoplanetary disks have been imaged at high spatial resolution with
the Atacama Large Millimeter/submillimeter Array (ALMA), revealing
physical structure that may indicate the presence of low-mass companions 
or planets in the disk, e.g., IRS 48, HD 142527, HD 100546, HL Tau, TW Hya, HD 97048, HD 163296, 
HD 169142, and AS 209 \citep{vanderMarel2013,Casassus2013,Walsh2014b,ALMApartnership2015,Andrews2016,Walsh2016,Isella2016,Fedele2017a,Fedele2017b}. 
Disks have a stratified vertical structure with an atomic upper layer, a chemically
active warm intermediate layer, and a dense, cold midplane. The
environment most conducive to planet formation is at the disk midplane,
where various molecular species such as H$_2$O, CO$_2$, CO, and N$_2$
freeze out onto dust grains, creating an icy mantle that enhances
grain sticking efficiency \citep{Bergin2007,Blum2008}. Molecular ices
can then be more easily incorporated into the bulk of planetary bodies
than their gas-phase counterparts. The location at which $\sim$50\% of a
given molecule has condensed into ices is called a snow line \change{or ice line}.

Probing the conditions of the midplane of the disk is difficult.
\change{Dust and molecular line opacities can obscure lower layers of the 
disk, particularly at radii close to the central star. Molecular snow lines 
can reside too near to the star to be directly observed, as is the case for H$_2$O 
\citep{Zhang2013,Piso2015,Banzatti2015}. They can also be obscured
by opacity effects in the outer disk, as is the case for the farther 
out CO snow line, where $^{12}$CO, $^{13}$CO, and even C$^{18}$O can 
remain optically thick at large radii \citep{Qi2015,Fedele2017a}.
\newchange{Direct determination of the CO snow line can be 
done using emission from the rarest CO isotopolgues
\citep[e.g.,][]{Yu2016,Zhang2017}, but only for the closest objects.}
To characterize the disk midplane environment, less abundant species must
be observed which trace chemical processes occurring deep in the disk, 
such as molecular freeze-out. DCO$^+$ has been suggested as an optically
thin molecular tracer of the midplane regions around the CO snow line 
and as a simultaneous tracer of ionization occurring in the intermediate
layers of the disk due to its formation via cold ($\lesssim$30 K) 
and warm ($\lesssim$100 K) deuterium fractionation 
pathways available in protoplanetary disks \citep{Mathews2013,Favre2015,Huang2017}.}

The disk around HD 169142 makes an excellent test-bed in which to 
explore the \newchange{chemistry of DCO$^+$ in protoplanetary disks and its 
usefulness as a tracer of disk midplane conditions}. \change{HD 169142 is one of a handful of disks
found to have millimeter dust rings, \newchange{and which also exhibits CO emission that extends} beyond the
edge of the millimeter grains \citep{ALMApartnership2015,Andrews2016,Isella2016,Walsh2016,Fedele2017a}.}
HD 169142 is an isolated system with a
Herbig Ae spectral type A8 Ve star and stellar mass
$M_{*}$ = 1.65 $M_{\odot}$ \citep{Grady2007,Blondel2006}. 
Recent distance measurements by Gaia
put the system at a distance of $d = 117 \pm 4$ pc \citep{Gaia2016}.
\newchange{The new distance results in a revised luminosity that is lower by 
a factor of $\sim$0.65, which places the age of the system closer to $\sim$10 Myr \citep{Pohl2017}.
The new age estimate is older than previous estimates of 6$_{-3}^{+6}$ Myr \citep{Grady2007}, 
but within the errors.}
With disk inclination $i$ = 13$^\circ$ and position angle 
P.A. = 5$^\circ$ \citep{Raman2006,Panic2008}, the system is viewed
close to face-on, allowing for accurate characterization of the
radial distribution of the continuum and molecular line emission.
\change{With an estimated total gas mass of $1.9 \times 10^{-2} \ M_{\odot}$
and $^{12}$CO extending out to $\sim$200 AU \citep{Fedele2017a}, 
the HD 169142 disk has a high concentration of gas.}
There is already known substructure in the dust around HD 169142.
\newchange{A hot inner ring of dust at $\sim$0.2 AU was detected
\citep{Wagner2015} within} a central dust cavity, and two dust 
\newchange{rings at $\sim$25 AU and $\sim$60 AU are clearly visible 
in the 1.3 millimeter continuum with ALMA \citep{Fedele2017a} and in 
scattered light with GPI and VLT/SPHERE \citep{Monnier2017,Pohl2017}.} The gap carved out 
between the rings may be indicative of ongoing planet formation. 
\newchange{An outer gap at $\sim$85 AU just beyond the edge
of the 1.3 millimeter continuum emission was also detected in scattered
light \citep{Pohl2017} and in 7 and 9 millimeter emission with the VLA \citep{Macias2017}.
While the millimeter grains terminate at $\sim$85 AU, the micron-sized
grains are present throughout radial extent of the gaseous disk.}
Determining the midplane conditions of this disk would provide insight into
the cold disk environment during the planet-building epoch.

This paper presents ALMA observations of the $J = 3-2$ transition
of DCO$^+$ toward HD 169142 and characterizes its distribution
throughout the disk. Section~\ref{sec:obs} describes the observations
and data reduction. The detection and distribution of DCO$^+$ throughout
the disk is detailed in Section~\ref{sec:res}. Modeling of the 
disk structure and DCO$^+$ emission is explained in Section~\ref{sec:mod}. 
Section~\ref{sec:disc} discusses the relationship between DCO$^+$
and the disk environment, followed by the conclusions in Section~\ref{sec:concl}.

\section{Observations and Reduction}
\label{sec:obs}

HD 169142 (J2000: R.A. = 18$^{\rm{h}}$24$^{\rm{m}}$29.776$^{\rm{s}}$, 
DEC = --29$^\circ$46$\arcmin$50.000$\arcsec$) was observed 
with ALMA in band 6 (211--275 GHz) with 35 antennas
on 2015 August 30 at a spatial resolution $\sim$0.3$\arcsec$. 
\change{The project code is ADS/JAO.ALMA\#2013.1.00592.S.
The data used in this work were reduced in the same
manner as \citet{Fedele2017a}. See their paper for further 
details on calibration, self-calibration, and continuum subtraction.}
Data reduction was performed with version 4.3.1 of the Common Astronomy 
Software Applications \citep[\textsc{casa};][]{McMullin2007}. 
Images were created using the \textsc{casa} task \textsc{clean},
with natural weighting for the lines \newchange{to enhance 
sensitivity}.

\begin{table}[!tp]
 \caption{HD 169142 Observational Parameters}
 \centering
 \label{tab:obs_par}
 \resizebox{8.5cm}{!}{
 \begin{tabular}{lc}
 \hline \hline
 \multicolumn{2}{c}{Project 2013.1.00592.S} \\
 \hline
 Date Observed & \multicolumn{1}{c}{2015 August 30} \\
 Baselines & \multicolumn{1}{c}{13 -- 1445 m | 10 -- 1120 k${\rm \lambda}$ } \\
 \hline
  & DCO$^+$ $J=3-2$ \\
 Rest frequency [GHz] & 216.11258 \\
 Synthesized beam [FWHM] & $0.37\arcsec \times 0.23\arcsec$ \\
 Position angle & --74.8$^\circ$ \\
 Channel width [km s$^{-1}$] & 0.085 \\
 rms noise [mJy beam$^{-1}$]  & 6 \\
 $v_{LSR}$ [km s$^{-1}$] & 6.9 \\
 FWHM [km s$^{-1}$] & 1.76 \\
 Integrated flux\tablefootmark{a} [mJy km s$^{-1}$]  & 730$\pm$73 \\
 Weighting & \multicolumn{1}{c}{natural} \\
 \hline
 \end{tabular}
 }
 \tablefoot{Flux calibration accuracy is taken to be 10\%. 
 \tablefoottext{a}{Line flux obtained after applying a Keplerian mask to the image cube (Section~\ref{sec:res}).}
 }
\end{table}

\change{The full data set contained observations of
the 1.3 mm continuum and the molecular lines $^{12}$CO $J=2-1$, $^{13}$CO $J=2-1$,
C$^{18}$O $J=2-1$, and DCO$^+$ $J=3-2$. The DCO$^+$ $J=3-2$
line at 216.1128 GHz was observed in the lower sideband and 
had a frequency (velocity) resolution of 61.0 kHz (0.084 km s$^{-1}$ ).
This work focuses on the analysis of the DCO$^+$ $J=3-2$ data 
and makes use of the 1.3 millimeter continuum and C$^{18}$O $J=2-1$ images.} 
Previous analysis of the continuum and the three CO
isotopologue lines was reported in \citet{Fedele2017a}. 
\citet{Macias2017} presented a brief analysis of the C$^{18}$O $J=2-1$ and DCO$^+$ $J=3-2$ data.
In their comparison of the molecular radial intensity profiles, the authors use a 
$uv$ taper to increase the signal-to-noise, resulting in a lower resolution DCO$^+$ 
image. We instead use a Keplerian mask to improve the signal-to-noise of the DCO$^+$
\newchange{integrated intensity} image, thus retaining the high spatial resolution. 
\newchange{We also present extensive modeling of the DCO$^+$ emission to explore the 
sensitivity of the emission to the disk physical conditions.} 
Table~\ref{tab:obs_par} summarizes the observational parameters
for the DCO$^+$ $J=3-2$ emission in this work.

%______________________________________________________________
\begin{figure*}[!tp]
 \centering
 \includegraphics[width=0.45\textwidth]{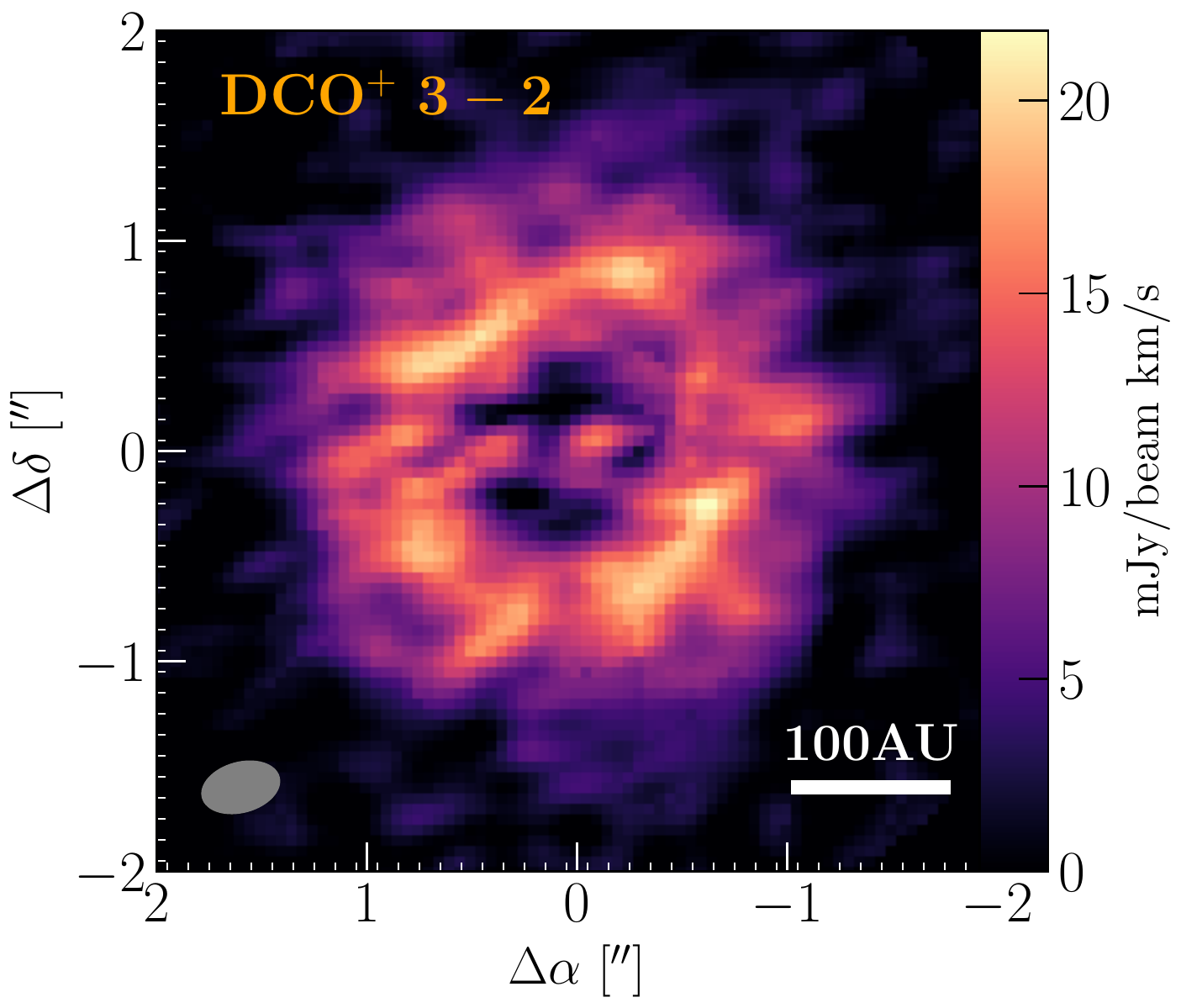}
 \hspace{1cm}
 \includegraphics[width=0.45\textwidth]{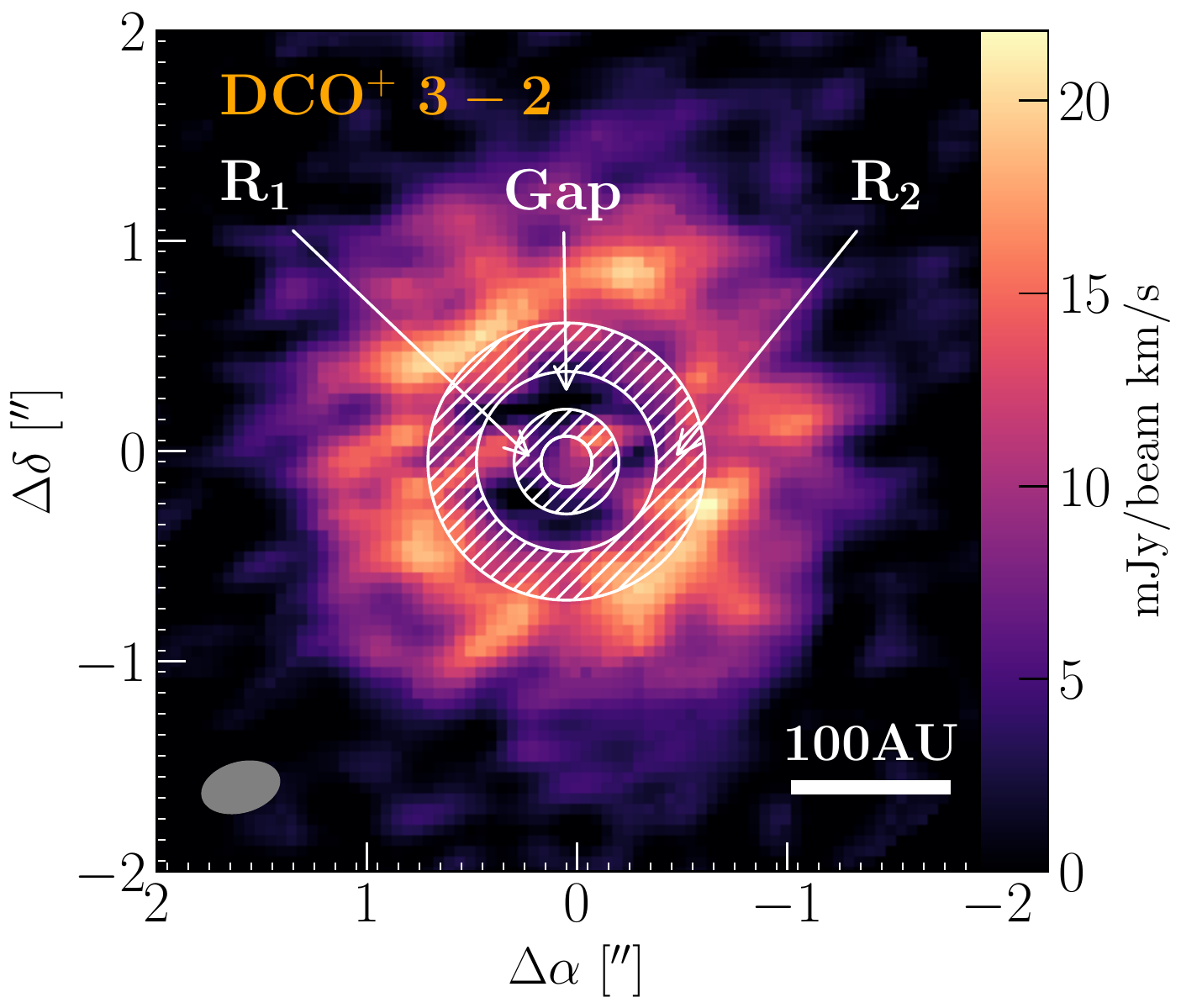} \\
 \vspace{0.5cm}
 \hspace{-0.7cm}
 \includegraphics[width=0.45\textwidth]{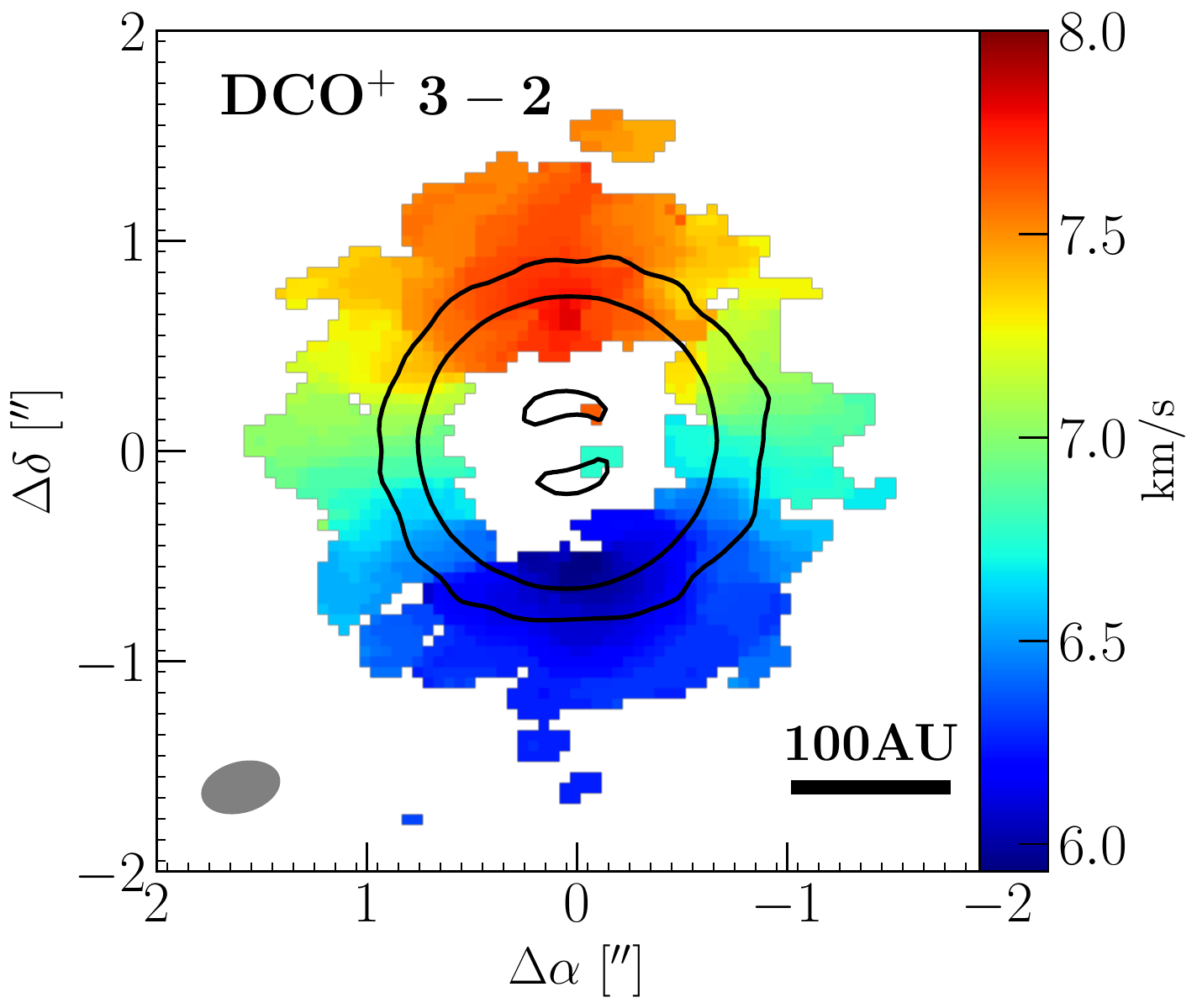}
 \hspace{0.5cm}
 \includegraphics[width=0.44\textwidth,height=0.38\textwidth]{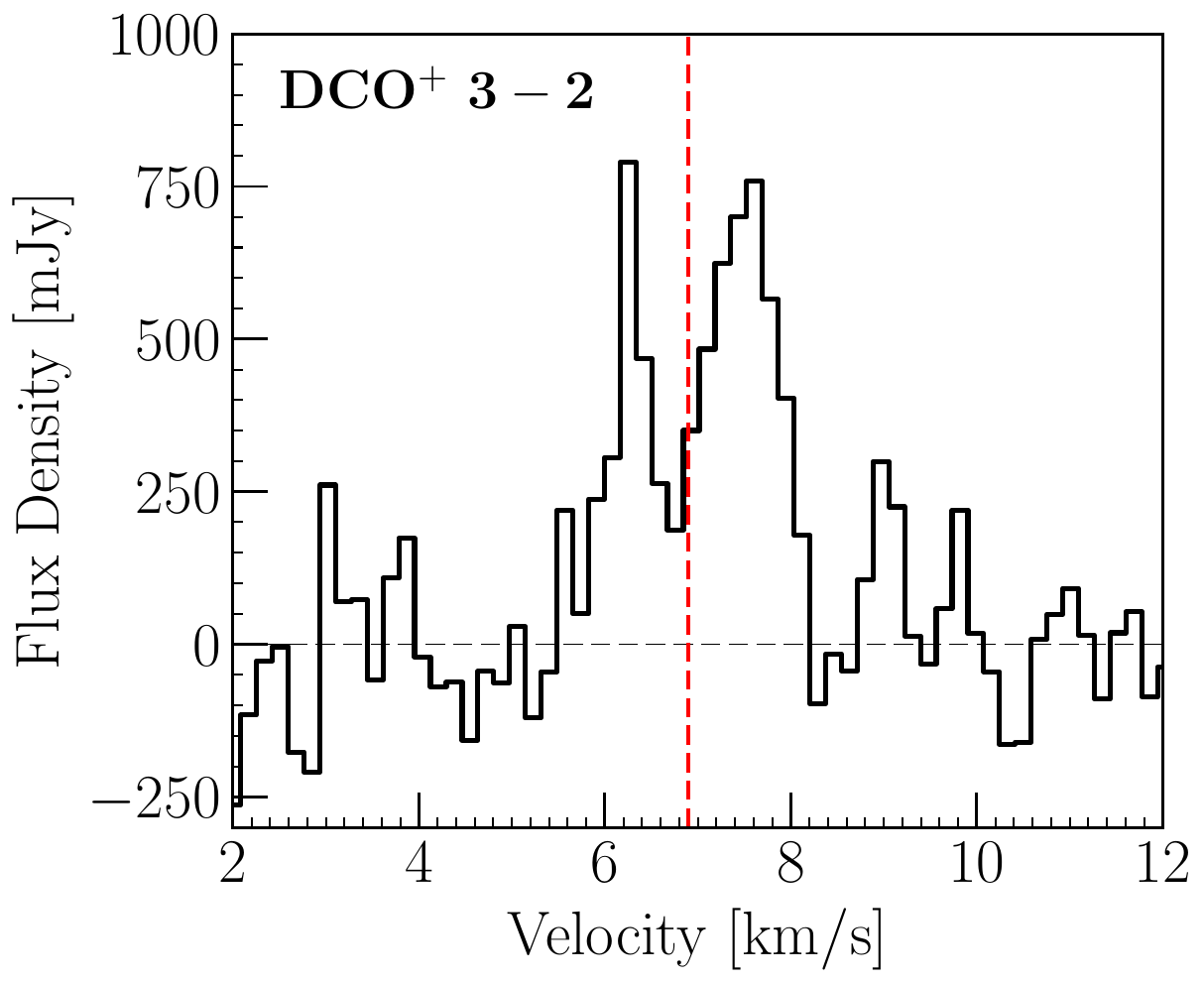}
 \caption{
 ({\it Top-left})
 Integrated intensity map of the DCO$^+$ $J=3-2$ line from 5.4 -- 8.8 km s$^{-1}$
 after applying a Keplerian mask to the image cube. 
 Synthesized beam and AU scale are shown in the lower corners.
 ({\it Top-right})
 \refchange{DCO$^+$ $J=3-2$ integrated intensity map overlaid with white marking the \newchange{model} inner dust
 cavity, the inner dust ring (R$_1$, hatched), the dust gap, and the outer dust ring (R$_2$, hatched).}
 Synthesized beam and AU scale are shown in the lower corners.
 ({\it Bottom-left}) Velocity-weighted coordinate map of the DCO$^+$ $J=3-2$ line,
 clipped at 3.5$\sigma$. Solid black contours show the 233 GHz/1.3 mm emission at
 7.0 $\times$ $10^{-5}$ Jy beam$^{-1}$ (1$\sigma$) $\times$ [5, 50, 200].
 Synthesized beam and AU scale are shown in the lower corners.
 ({\it Bottom-right)} Disk-integrated spectrum of the DCO$^{+}$ $J=3-2$ line
 before Keplerian masking, Hanning smoothed to 0.17 km s$^{-1}$ channels.
 The horizontal dashed black line indicates the \newchange{continuum-subtracted spectral} baseline.
 The vertical red line shows the systemic velocity at 6.9 km s$^{-1}$.
 }
 \label{fig:spec_mom_maps}
\end{figure*}

\section{Results}
\label{sec:res}

The DCO$^+$ $J=3-2$ line in the disk around HD 169142 was readily detected and imaged at 
$0.37\arcsec \times 0.23\arcsec$ [$43\times27$ AU at 117 pc] spatial resolution, with 
beam P.A. = --74.8$^{\circ}$. The systemic velocity is 6.9 km s$^{-1}$ \citep{Fedele2017a}.
The spectrum shown in Figure~\ref{fig:spec_mom_maps} 
was extracted from the original self-calibrated,
continuum-subtracted \textsc{clean} image.
The right ascension and declination axes of the image cube
are collapsed over a circular region with 
radius 1.75$\arcsec$ centered on the source position. 

To enhance the signal-to-noise of the DCO$^+$ emission maps
and radial profile, a mask in right ascension, declination,
and velocity was applied to the original image cube data,
following \citet{Carney2017} and \citet{Salinas2017}.
The mask is based on the \newchange{velocity profile of a rotating disk}, 
which is assumed to be Keplerian \newchange{around a central stellar mass} of 
$M = 1.65 \ M_{\odot}$ \citep{Blondel2006}. 
A subset of pixels are identified in each velocity channel
where the pixel Keplerian velocity matches the
Doppler-shifted line velocity. Pixels with velocities that do not
match the Keplerian rotational profile criteria are masked.
Appendix~\ref{app:B} shows the DCO$^+$ $J=3-2$ channel maps with the
Keplerian mask outline visible as the blue contours.
To obtain the integrated line flux for DCO$^+$ $J=3-2$ 
reported in Table~\ref{tab:obs_par}, the spectrum was extracted from 
the image cube after applying the Keplerian mask and
integrated over the velocity range 5.4 -- 8.8 km s$^{-1}$.

\subsection{Radial distribution of DCO$^+$}
\label{sec:res_distr}

\begin{figure}[!tp]
 \centering
 \includegraphics[width=0.45\textwidth]{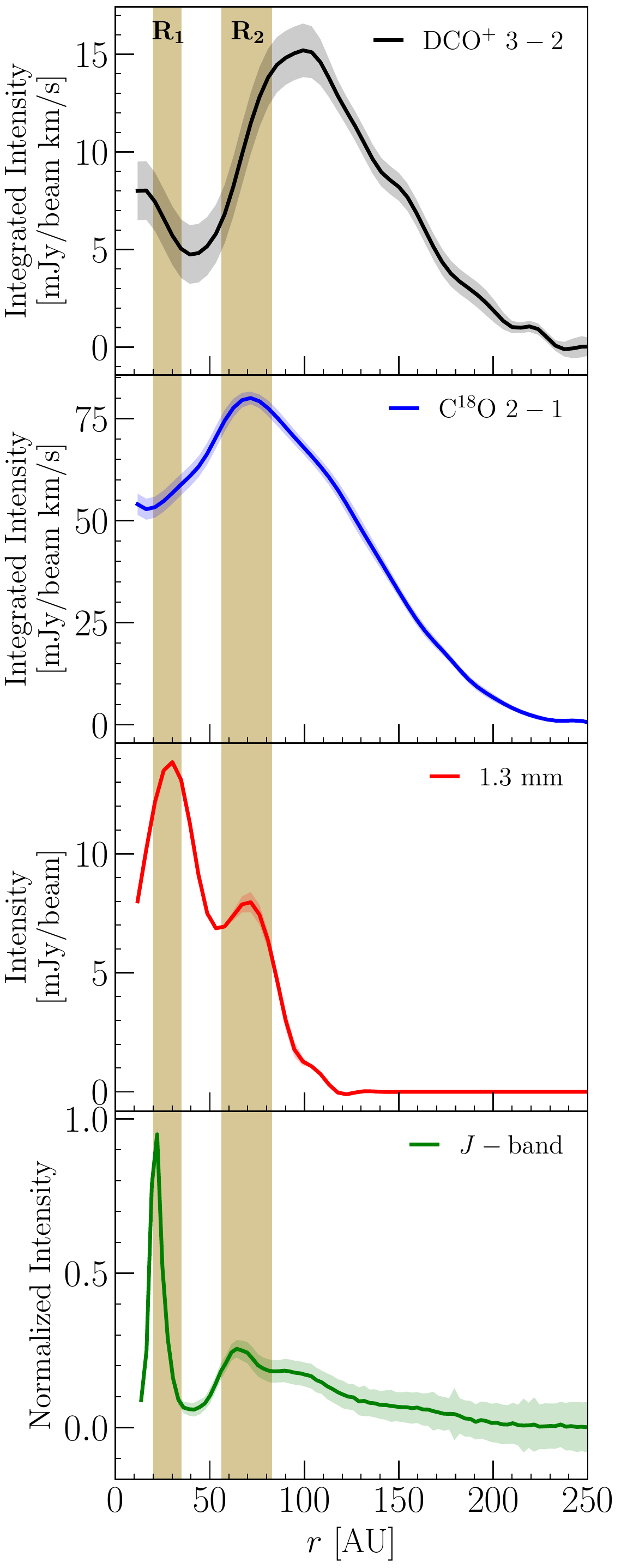}
 \caption{Azimuthally-averaged radial intensity profiles of DCO$^+$ ({\it top}),
 C$^{18}$O ({\it top-middle}), the 1.3 millimeter continuum ({\it bottom-middle}),
 and the $J$-band ($\sim$1.2 $\mu$m) polarized light ({\it bottom}).
 Shaded regions represent 1$\sigma$ errors \newchange{on the intensity}.
 }
 \label{fig:radint_dcop_c18o_cont}
\end{figure}

The DCO$^+$ emission has a ring-like morphology in this disk,
with the majority of emission originating in a region between 
$0.4\arcsec$--$1.4\arcsec$ [47--164 AU at 117 pc],
based on the velocity-weighted coordinate (first-order moment) map in
Figure~\ref{fig:spec_mom_maps} obtained from applying a 3.5$\sigma$ clip 
to the emission in the Keplerian-masked DCO$^+$ image cube. The ring
extends significantly beyond the outer edge of the 1.3 millimeter continuum, similar
to the $^{12}$CO, $^{13}$CO, and C$^{18}$O molecular lines \citep{Fedele2017a}.
Figure~\ref{fig:spec_mom_maps} also shows the integrated intensity (zero-order moment)
map from velocity channels 5.4 -- 8.8 km s$^{-1}$. 
\newchange{Applying the Keplerian mask improved the signal-to-noise
ratio of the integrated intensity image by a factor of three, from 5 to 15.}
The radial profile for DCO$^+$
in Figure~\ref{fig:radint_dcop_c18o_cont}
is obtained by taking the mean intensity in azimuthally-averaged elliptical annuli 
projected to an inclination $i = 13^{\circ}$ and position angle ${\rm P.A. = 5^{\circ}}$.
The radial bin size was set to 0.1$\arcsec$ [11.7 AU at 117 pc].
\change{Errors are calculated as the standard deviation of the pixel intensity
contained within each annulus divided by the square root of the number of beams.}

With the increased signal-to-noise of the averaged annuli, 
it is clear from the radial intensity profile that
DCO$^+$ extends out to $\sim$230 AU and peaks at a radius of $\sim$100 AU. 
Within 100 AU, there is a gap
between $\sim$30--60 AU where the intensity drops, with some
DCO$^+$ emission returning at radii $\lesssim$30 AU.
Given the errors on the curve, the actual
drop in emission in the 30--60 AU region may be \newchange{small} (see Figure~\ref{fig:radint_dcop_c18o_cont}).
\change{At $\sim$150 AU there is a knee in the radial profile,
and at $\sim$200 AU there is a distinct bump.}

The intensity at $r \lesssim 30$ AU already suggests that 
there is a warm component to the DCO$^+$ emission, as 
temperatures in this region of the disk are \change{too high to allow
the cold deuterium fractionation pathway to be active.} As seen in Figure~\ref{fig:radint_dcop_c18o_cont},
the dip in DCO$^+$ intensity from $r$ = 30--60 AU corresponds well
to the gap between the two dust rings, indicated by the filled regions.
With less dust and gas in the gap, the overall surface density
profile falls dramatically, causing a corresponding dip in the DCO$^+$ radial profile,
\newchange{more prominently than seen in C$^{18}$O.} 

The DCO$^+$ intensity increases significantly within the outer dust ring, 
which is to be expected if DCO$^+$ is forming near the midplane where the dust
temperature is sufficiently low for some degree of CO freeze-out. Interestingly,
the peak in the radial profile at $\sim$100 AU is beyond the outer edge of the second 
dust ring, and emission is present throughout the outer disk.
This \change{suggests} that beyond the 1.3 mm continuum the disk remains cold
due to the presence of micron-sized dust grains, \newchange{as observed by 
\citet{Quanz2013}, \citet{Monnier2017}, and \citet{Pohl2017}.}

\change{In addition, Figure~\ref{fig:radint_dcop_c18o_cont} compares
the radial profiles of DCO$^+$, C$^{18}$O, the 1.3 millimeter continuum,
\newchange{and the $J$-band (1.2 $\mu$m) polarized intensity. The polarized intensity data is from VLT/SPHERE, 
with the radial profile obtained after azimuthally averaging the deprojection of the $r^2$-scaled $J$-band
$Q_\phi$ image and normalizing to the maximum brightness of the inner ring \citep{Pohl2017}.
The dust rings R1 and R2 are clearly visible in both the millimeter
emission from ALMA and the micron emission from VLT/SPHERE. The $J$-band profile shows small
grains existing throughout the extent of the gaseous disk out to $\sim$200 AU.}
The radial profiles show that the DCO$^+$ emission is highly sensitive to changes
in the disk structure, whereas C$^{18}$O is less affected.
The feature at $\sim$200 AU reveals that there is
some mechanism in the disk causing more DCO$^+$ emission than would
be expected for a smoothly decreasing abundance. There may be an 
accompanying slope change of the C$^{18}$O at $\sim$190 AU, but 
it is difficult to discern in Figure~\ref{fig:radint_dcop_c18o_cont}. 
A feature at large radii in the C$^{18}$O is more apparent in the radial 
slice along the major axis of the disk shown in Figure 3 of \citet{Fedele2017a},
where a small bump can be seen at $\sim$1.7$\arcsec$ [200 AU at 117 pc],
hinting at outer disk structure in C$^{18}$O.
At these radii the DCO$^+$ is tracing midplane  
substructure in the disk that is not as apparent in the more 
abundant, optically thick CO isotopologues.}

\subsection{Column density and disk-averaged abundance in LTE}
\label{sec:res_abun}

\begin{table}[!tp]
 \caption{DCO$^+$ disk-averaged column density and abundance.}
 \centering
 \label{tab:col_dens_abun}
 \begin{tabular}{lccc}
 \hline \hline
 $T_{\rm ex}$ & $N_{\rm avg}$ & $M_{\rm disk}$ & $N({\rm DCO^+})/N({\rm H_2})$ \\
 $ $ [K] & [cm$^{-2}$] & [M$_\odot$] \\
 \hline
 25 & $3.7 \times 10^{11}$ & $1.9 \times 10^{-2}$ & $9.0 \times 10^{-13}$ \\
 50 & $4.8 \times 10^{11}$ & $1.9 \times 10^{-2}$ & $1.2 \times 10^{-12}$ \\
 75 & $6.3 \times 10^{11}$ & $1.9 \times 10^{-2}$ & $1.5 \times 10^{-12}$ \\
 \hline
 \end{tabular}
\end{table}

\newchange{We} estimated the disk-averaged abundance
of the observed DCO$^+$ based on the total integrated line flux, an assumed excitation 
temperature, and the total disk mass. Following the formula used by 
\citet{Remijan2003} and \citet{Miao1995} for optically thin emission
in local thermodynamic equilibrium (LTE), we can estimate the column density

\begin{equation}
 \label{eq:col_dens}
 N = 2.04 \frac{\int I_{\nu} dv}{\theta_{\rm a} \theta_{\rm b}} \frac{Q_{\rm rot} \ {\rm exp} (E_{\rm u} / T_{\rm ex})}{\nu^3 \langle S_{\rm ij} \mu^2 \rangle}
 \times 10^{20} \ {\rm cm}^{-2},
\end{equation}

\noindent where $\int I_{\nu} dv$ is the integrated line flux in Jy beam$^{-1}$ km s$^{-1}$,
$\theta_{\rm a}$ and $\theta_{\rm b}$ correspond to the
semi-major and semi-minor axes of the synthesized beam in arcseconds,  
$T_{\rm ex}$ is the excitation temperature in K, and
$\nu$ is the rest frequency of the transition in GHz. 
The partition function ($Q_{\rm rot}$), upper energy level 
($E_{\rm u}$, in K), and the temperature-independent
transition strength and dipole moment ($S_{\rm ij} \mu^2$, in debye$^2$)
for the DCO$^+$ molecule are taken from the CDMS database
\citep{Muller2005}.

DCO$^+$ is expected to form primarily in the midplane
close to the CO freeze-out temperature, where gas densities
are typically higher \citep[$\sim$10$^9$ cm$^{-3}$;][]{Walsh2014a}
than the critical density of the $J = 3-2$ transition at 20--30 K
\citep[$\sim$2$\times$10$^6$ cm$^{-3}$;][]{Flower1999}. 
Under these conditions, LTE is a reasonable assumption.
\newchange{Furthermore, the density of the H$_2$ gas taken
from the \citet{Fedele2017a} model (see 
Figure~\ref{fig:mod_dens_temp_co}) is greater than the critical 
density of DCO$^+$ for $z/r < 0.3$. Therefore, only if DCO$^+$
is present solely in the diffuse upper disk layers where the gas
and dust temperature have decoupled would LTE be an unreasonable assumption.} Currently, formation
routes for DCO$^+$ place the molecule in significant abundance
only in the intermediate disk layers ($\lesssim$100 K) and near the 
midplane, further justifying the use of LTE.

\newchange{We explore excitation temperatures of 25, 50, and 75 K,
which cover the range of expected DCO$^+$ emitting regions \citep{Mathews2013,Favre2015}.}
The total integrated line flux and excitation temperature are used to 
calculate a disk-averaged DCO$^+$ column density.
\newchange{Assuming optically thin emission, the disk-averaged column density was then}
used to estimate the total number of DCO$^+$ molecules
in the disk, $N({\rm DCO^+}) = N_{\rm avg} \times (a \times b)$,
where $(a \times b)$ is the total emitting area of DCO$^+$.
Assuming the total disk mass is primarily molecular hydrogen,
we can estimate the total number of H$_2$ molecules,
$N({\rm H_2}) = M_{\rm disk} / m_{\rm H_2}$, where $m_{\rm H_2}$
is the molecular hydrogen mass.
The emitting area is set to $a = b = 3\arcsec$ based on the
\change{diameter} of emission in the integrated intensity map, and the total
disk mass is $1.9 \times 10^{-2} \ M_{\odot}$.
Table~\ref{tab:col_dens_abun} shows the disk-averaged column 
density and abundance \newchange{for $T_{\rm ex}$ = 25, 50, and 75 K, 
which are consistent to within a factor of two over the temperature range. 
DCO$^{+}$ column densities of order $10^{11} - 10^{12}$ are similar to the 
values reported for HD 163296 \citep{Mathews2013,Salinas2017}, 
TW Hya \citep{Qi2008}, and DM Tau \citep{Teague2015}.}

\section{Modeling DCO$^+$ emission}
\label{sec:mod}

\begin{figure*}[!t]
 \centering
 \includegraphics[height=6.3cm]{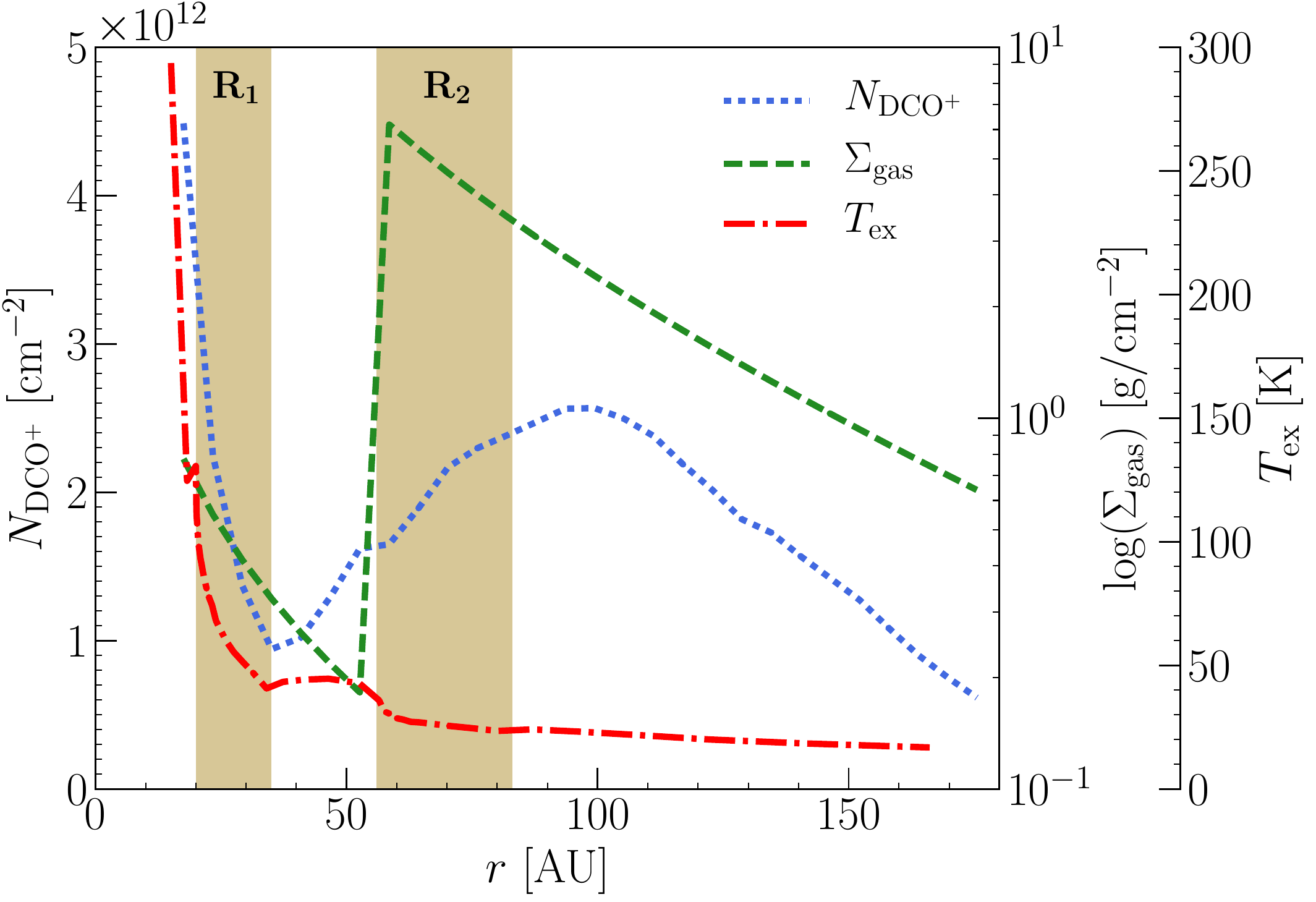} 
 \hspace{0.5cm}
 \includegraphics[height=6.1cm]{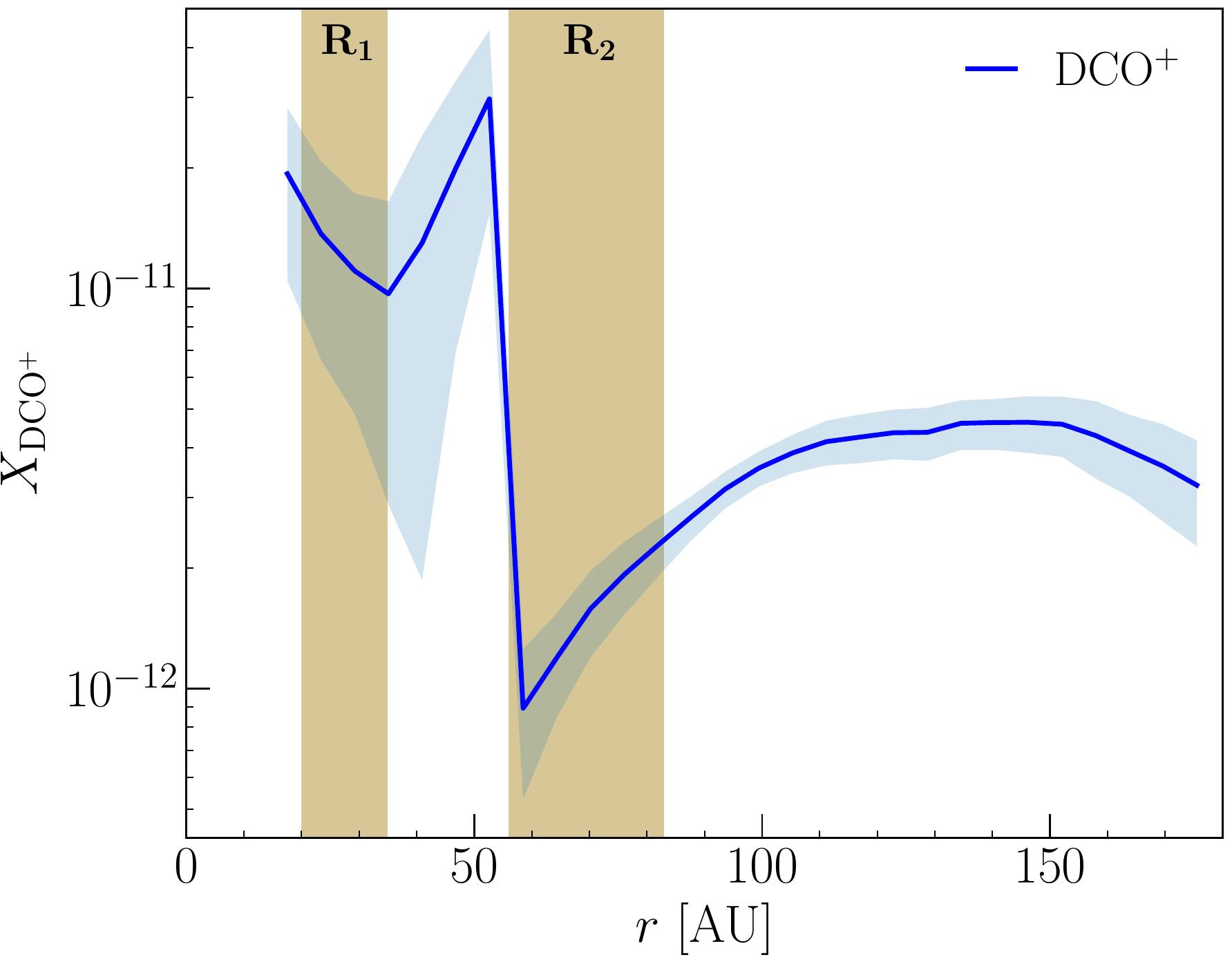}
 \caption{\newchange{One-dimensional, radial structure in the 
 HD 169142 disk.}
 Vertical brown shaded regions represent the dust rings.
 ({\it Left}) The DCO$^{+}$ radial column density	
 (dotted blue) calculated from Equation~\ref{eq:col_dens} using
 \newchange{the DCO$^{+}$ the radial intensity profile (see 
 Figure~\ref{fig:radint_dcop_c18o_cont}) and the midplane gas
 temperature of the optimized \textsc{dali} model
 (dot-dash red) as the $T_{\rm ex}$ profile. 
 The gas surface density profile (dashed green) 
 was used to derive the DCO$^{+}$ radial abundance.}
 ({\it Right}) Radial abundance structure of DCO$^{+}$ with respect to H$_2$. 
 The blue shaded region represents 1$\sigma$ errors \newchange{on the abundance}. 
 }
 \label{fig:dcop_radabun}
\end{figure*}

The aim of modeling the DCO$^+$ emission in this disk was to
determine the \newchange{midplane conditions which create sufficient
production of DCO$^+$ in the outer disk, and to estimate the contribution of 
cold and warm formation routes} to the overall DCO$^+$ abundance. The initial physical 
structure of the HD 169142 disk is adopted from \citet{Fedele2017a}, 
who constrained the density and temperature structure by simultaneously fitting
the radial distribution of the 1.3 mm continuum and three 
CO isotopologues: $^{12}$CO $J=2-1$, $^{13}$CO $J=2-1$, and C$^{18}$O $J=2-1$.
\newchange{The disk structure is then optimized to include a small
grain dust population throughout the disk that was absent in the original
model. With the optimized disk structure, we then reproduce the DCO$^+$ 
radial intensity profile in a parameterized way with a simple deuterium chemical network.}

\subsection{Fiducial physical structure}
\label{sec:mod_fid_phys}

We use the thermo-chemical code \underline{D}ust \underline{A}nd 
\underline{LI}nes \citep[\textsc{dali};][]{Bruderer2012,Bruderer2013}
to \newchange{obtain the physical disk structure}. Input for \textsc{dali} consists of
a blackbody radiation field with $T_{\rm eff}$ =  8400 K to estimate 
the stellar photosphere and a power-law gas surface density with an exponential drop-off

\begin{equation}
  \label{eq:surf_dens}
  \Sigma_{\rm gas} = \Sigma_{\rm c} \left( \frac{R}{R_{\rm c}} \right) ^{-\gamma} \left[ {\rm exp} - \left( \frac{R}{R_{\rm c}} \right) ^{2-\gamma} \right],
\end{equation}

\noindent where $R_{\rm c}$ (100 AU) is the critical radius, $\Sigma_{\rm c}$
(6.5 g cm$^{-2}$) is the value of the gas surface density at the critical radius
\newchange{and $\gamma$ (1.0) is the power-law exponent}. 
The \newrefchange{initial} dust surface density is extrapolated from the gas surface density by 
assuming a gas-to-dust ratio ($\Delta_{\rm gd} = 80$) such that 
$\Sigma_{\rm dust} = \Sigma_{\rm gas} / \Delta_{\rm gd}$.
The vertical gas density is described by a Gaussian distribution with 
a scale height $h = h_{\rm c}(R/R_{\rm c})^{\psi}$ that depends on the 
disk radius and the flaring exponent $\psi$ (0.0) \newchange{with a critical 
scale height, $h_{\rm c}$ (0.07), defined at the critical radius}.

Dust settling is approximated in \textsc{dali} by considering two different
populations of dust grains following the power-law description from \citet{DAlessio2006},
with a power-law exponent $p$ = 3.5. The small grains (0.005 -- 1 $\mu$m) have a scale
height $h$ while large grains (0.005 -- 1000 $\mu$m) have a scale height $h\chi$, where
\newchange{the settling parameter $\chi$ (0.2) is in the range 0--1}. 
The fractional distribution between the 
two populations of dust grains is set by the parameter $f_{\rm large}$ (0.85), which 
results in dust surface densities of $\Sigma_{\rm dust} f_{\rm large}$ for large 
grains and $\Sigma_{\rm dust} (1 - f_{\rm large})$ for the small grains.

\textsc{dali} solves for dust temperatures
and radiation field strength in each grid cell using 2D radiative transfer, then
determines the heating-cooling balance of the gas, molecular excitation, and 
chemical abundances based on an input chemical network.
\newchange{The \textsc{dali} model described in this work uses the ISO chemical network, 
which includes CO freeze-out and CO isotope-selective photodissociation \citep{Miotello2014,Miotello2016}.}
% \change{Finally, \textsc{dali} uses a ray-tracing algorithm to solve 
% for molecular radiative transfer and outputs synthetic spectral image cubes.}

\citet{Fedele2017a} modified the surface density profile from Equation~\ref{eq:surf_dens} to
include \newchange{gas depletion in the inner disk and} \newrefchange{two millimeter dust rings: 
R$_1$ from 20 -- 35 AU and R$_2$ from 56 -- 83 AU. These modifications result in a radially variable gas-to-dust ratio 
throughout the disk, with $\Delta_{\rm gd} = 80$ valid only for the R$_2$ outer dust ring 
from 56 -- 83 AU.} 
For a full description and table of values including the parameter ranges and 
best-fit parameters of their fiducial \textsc{dali} model, see Section 5.2, Figure 5, 
and Table 2 of their paper. \newchange{Their fiducial model only
included a small grain dust population within the millimeter rings since they fit only
the millimeter emission. We expanded on their fiducial model by including small grains 
in other regions of the disk and optimized the model parameters to keep the fit to the spectral
energy distribution, 1.3 mm emission, and CO isotopologues. See Appendix~\ref{app:A} for details.}

\begin{figure}[!bp]
 \centering
 \includegraphics[width=0.44\textwidth]{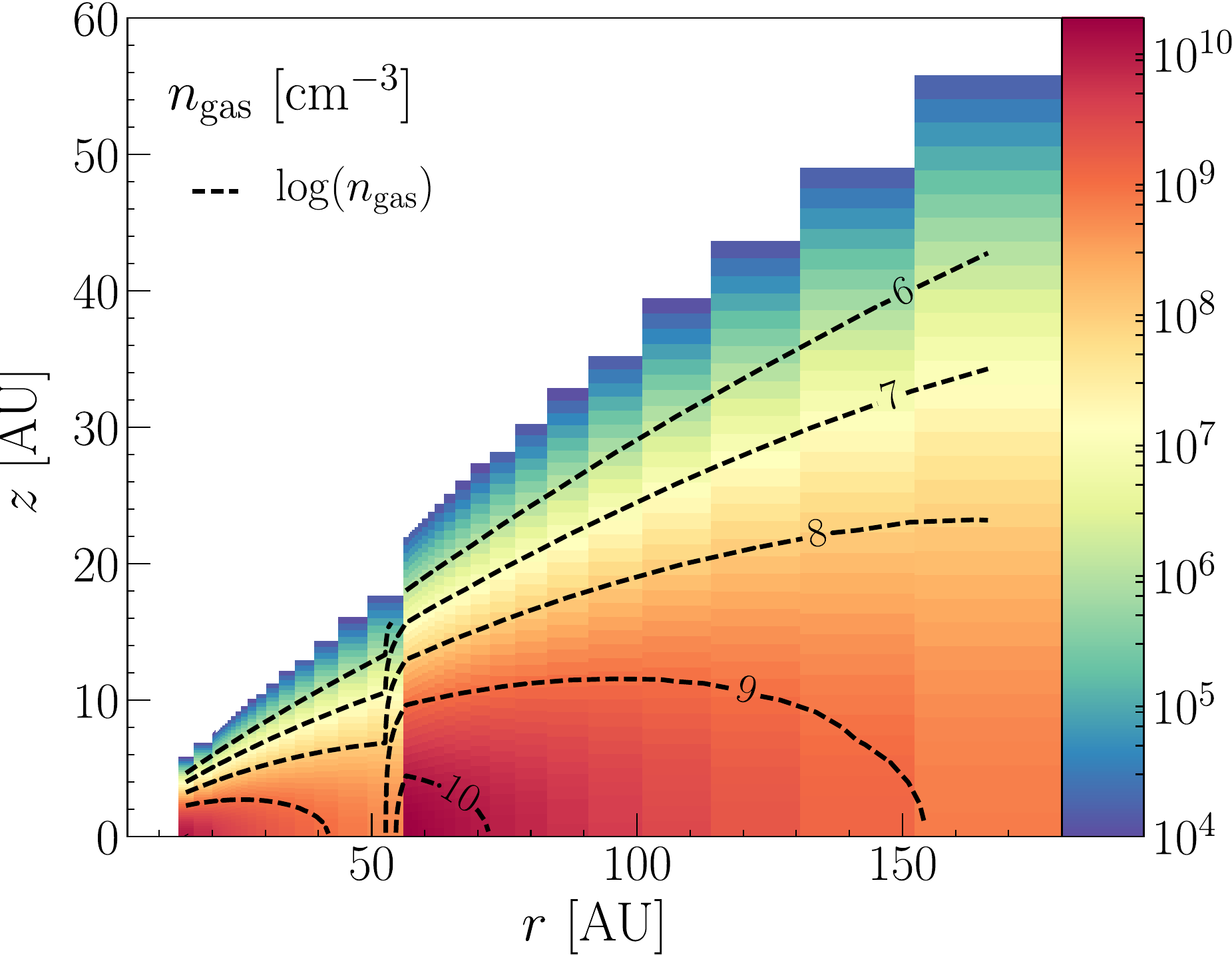}  \\
 \hspace{-0.3cm} 
 \includegraphics[width=0.425\textwidth]{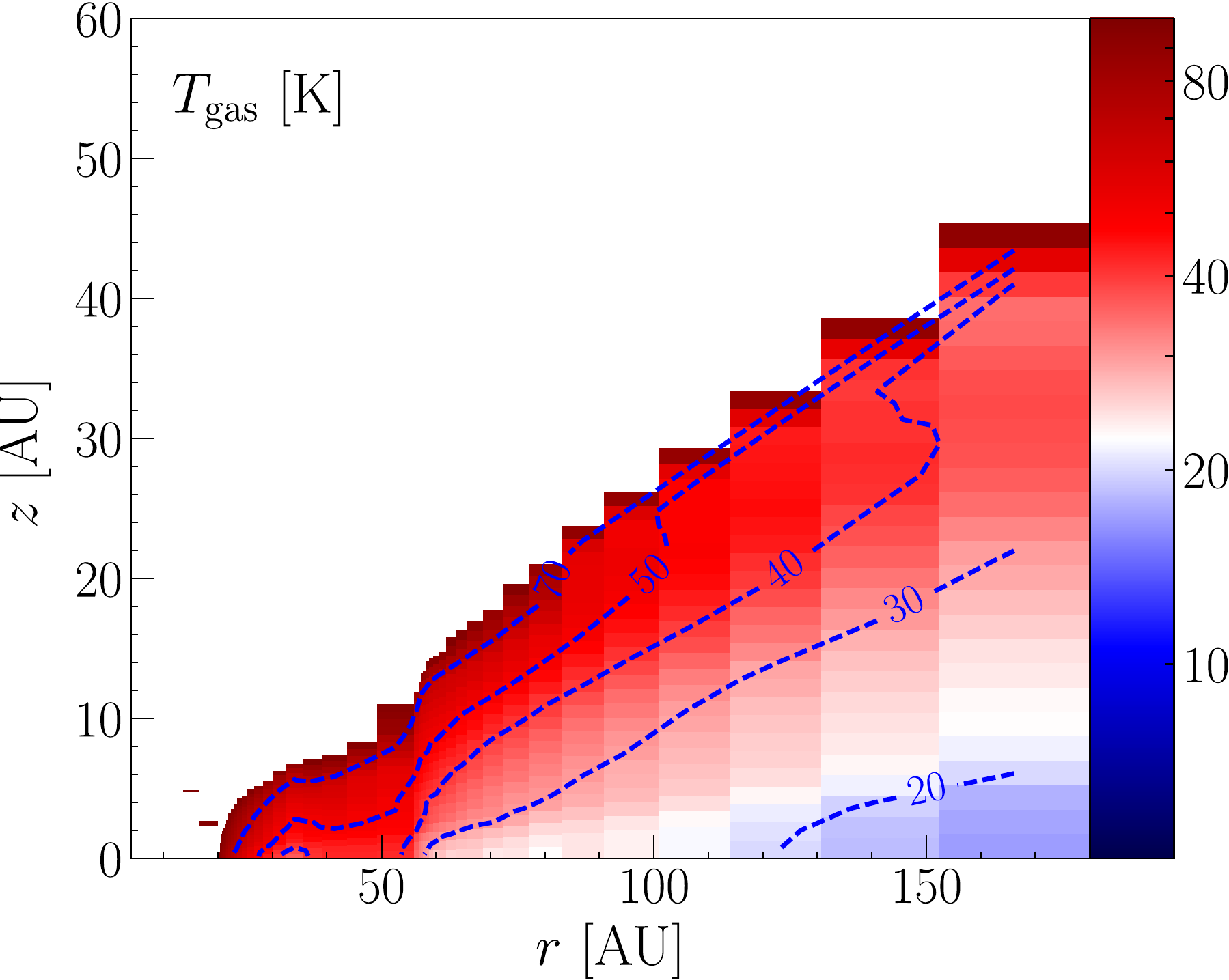}  \\
 \hspace{1.1cm}
 \includegraphics[width=0.45\textwidth]{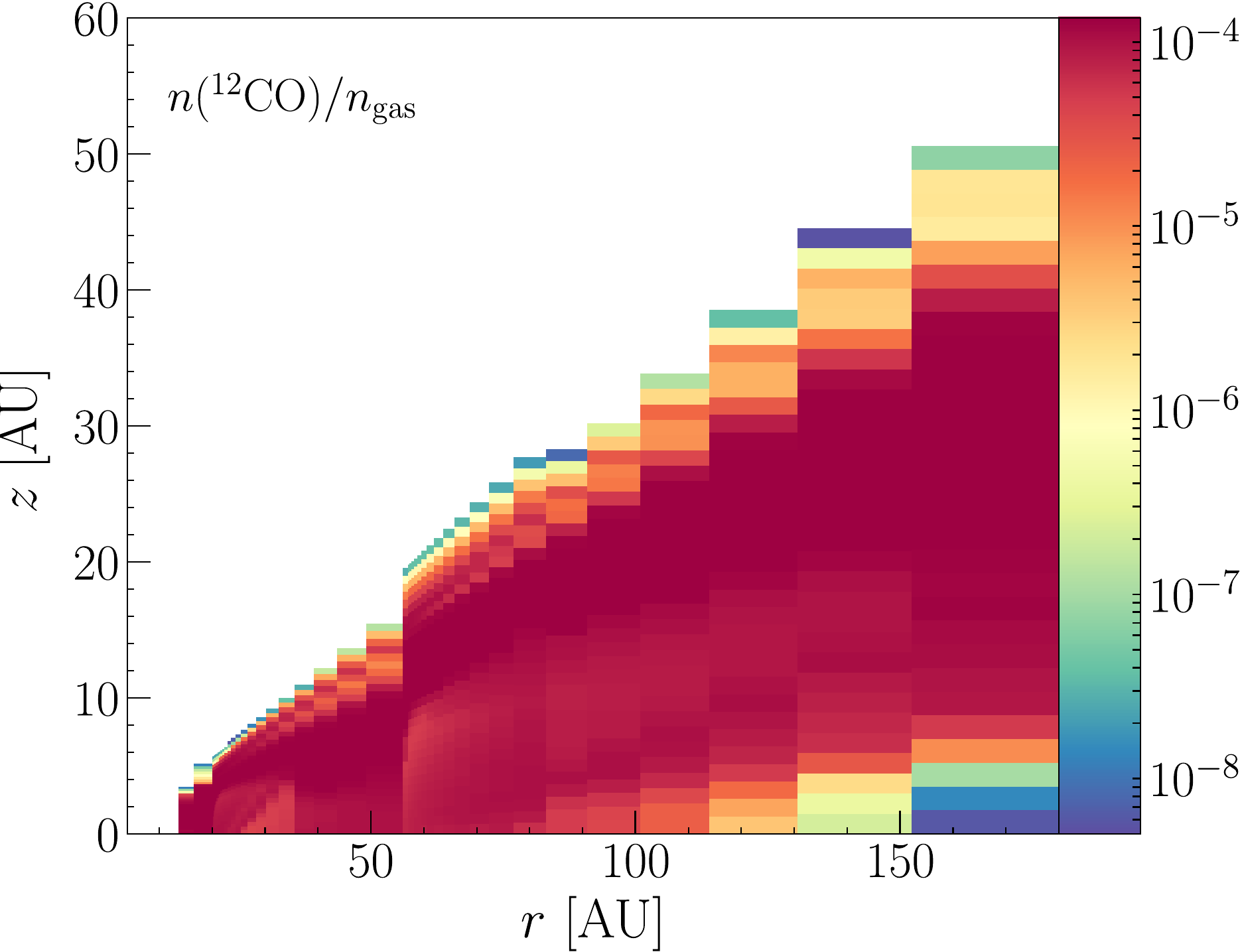}
 \caption{\newchange{Two-dimensional physical structure of the 
 HD 169142 disk} from the optimized \textsc{dali} model (\newrefchange{Appendix~\ref{app:A}}).
 ({\it Top}) Gas density structure. The gas density contours (dashed black)
 are shown as log($n_{\rm gas}$)
 ({\it Middle}) Gas temperature structure below 100 K.
 \refchange{Temperature contours are shown in dashed blue.}
 ({\it Bottom}) CO abundance structure with respect to H$_2$.
 }
 \label{fig:mod_dens_temp_co}
\end{figure}

\subsection{Vertically-averaged radial abundance profile in LTE}
\label{sec:mod_rad_abun}

A radial abundance profile for the observed DCO$^{+}$ emission
can be calculated using the method outlined in Section~\ref{sec:res_abun}.
\newchange{Rather than estimate a global excitation temperature for DCO$^{+}$,
with the model physical structure outlined in the previous section
we can now obtain a radial description of the excitation temperature.}
To estimate $T_{\rm ex}$, which will be equivalent to the kinetic gas
temperature assuming LTE, the midplane gas temperature in each radial bin
was taken from the optimized \textsc{dali} model.
The line intensity was extracted from the integrated intensity map 
in radial bins as in Section~\ref{sec:res_distr},
with a bin size of 0.1$\arcsec$ [11.7 AU].
Equation~\ref{eq:col_dens} was then used to determine the radial
column density of DCO$^{+}$. Assuming the gas is composed primarily of 
H$_2$, the H$_2$ mass can be used to convert the gas surface density profile
of the \textsc{dali} model into a gas column density profile. 
Dividing the DCO$^{+}$ radial column density by the gas radial column density 
gives a radial abundance profile for DCO$^{+}$.

Figure~\ref{fig:dcop_radabun} shows the DCO$^{+}$ radial abundance.
The profile has an inner radius of 13 AU and an outer radius
at 180 AU, corresponding to the inner and outer radii of the model gas surface
density. \newchange{Beyond $r \sim 50$ AU, the abundance increases with
radius by a factor of about 5 with values ranging from 
$1-5 \times 10^{-12}$, which are comparable values to the DCO$^{+}$ radial 
abundance estimates for HD 163296 \citep{Salinas2017}. Similar trends
of DCO$^{+}$ abundance increasing with radius have been observed 
in TW Hya and DM Tau \citep{Qi2008,Teague2015}.}
The sharp increase in abundance at $r \sim 50$ AU is due to the 
$\delta_{\rm gas, gap} = 0.025$ depletion factor in the surface
density of the gas for $r < R_{\rm gap \ out}$ (56 AU). 
\change{Within the errors, the abundance profile remains relatively
flat at radii less than $R_{\rm gap \ out}$. Due to the proximity
to the central star, DCO$^{+}$ in this region is likely formed via
the warm deuterium fractionation pathway.}

\subsection{Parameterized models}
\label{sec:mod_colddisk}

\newchange{We move from the one-dimensional derivation
of the DCO$^+$ radial profiles to a two-dimensional
DCO$^+$ structure to explore variations in abundance for
different radii and heights in the disk. For this we parameterize 
the HD 169142 physical disk structure obtained from the optimized 
\textsc{dali} model, \newrefchange{shown in Figure~\ref{fig:mod_dens_temp_co}.}} 
\textsc{dali} calculates the dust temperature, 
the local radiation field, heating and cooling rates,
molecular abundances, and gas temperature self-consistently, making it difficult to
isolate and explore individual parameters that may affect DCO$^+$ emission.
We examine the effect of alterations to the disk gas temperature and CO abundance on 
DCO$^+$ production by employing a simple, parameterized modeling technique
using the \refchange{steady-state, analytic} chemical code (hereafter \textsc{dco+ chemnet}) 
from \citet{Murillo2015}. \refchange{This time-independent chemical model is preferred 
because the chemical timescales for gas-phase reactions are sufficiently fast 
that a steady state is achieved at times much shorter than the expected lifetime of 
the HD 169142 disk ($\sim$10 Myr). \citet{Murillo2015} have already shown that the \textsc{dco+ chemnet}
code reproduces the trends of full, time-dependent chemical models for protostellar envelopes.
We extend this treatment to the protoplanetary disk environment.}
See Section 4.1, Table 1, and subsection 4.2.3 in their paper
for a full description of the chemical network, the simplified set of
DCO$^+$ formation and destruction reactions via the cold deuterium fractionation pathway, 
and a comparison to a full chemical network.

The \textsc{dco+ chemnet} code takes as input the gas density, gas temperature, and CO abundance 
structure. The input profiles are taken from the output 
of the optimized \textsc{dali} model described in \newrefchange{Appendix~\ref{app:A}}.
\newchange{Because \textsc{dco+ chemnet} is a single-point code, each grid
cell in ($r$, $z$) of the \textsc{dali} disk grid was run separately} to capture the full 2D model structure. 
The gas density of the parameterized models is assumed equal to the molecular hydrogen density, $n_{\rm H_2}$, and 
the HD abundance is constant at $X_{\rm HD} = n_{\rm HD}/n_{\rm H} = 10^{-5}$ throughout the disk. 
\refchange{The cosmic ray ionization rate of H$_2$ is set to $\zeta_{\rm cr} = 1.26 \times 10^{-17}$ s$^{-1}$.}

\change{The parameterized models are used to probe the conditions near the midplane
of the outer disk, a region \newchange{to which previously observed CO isotopologue}
tracers are not sensitive. The CO observations do
not probe all the way down to the midplane in the outer disk and
only a small amount of CO is absent from the total column density}, while 
DCO$^+$ production is highly sensitive to the midplane conditions
\citep{Mathews2013,Qi2015,Favre2015,Huang2017}. 
The CO lines become optically thick beyond $R_{\rm dust \ out}$ (83 AU) near the midplane
\newchange{of the optimized \textsc{dali} model}
($z/r$ < 0.2 and < 0.1 for $^{12}$CO and C$^{18}$O, respectively)
so the physical structure is not \newchange{constrained} in this region. 
Therefore, in all parameterized models, only the region $r > 83$ AU and $z/r$ < 0.1 was altered since
this is the $\tau_{\rm mm}$ = 1 surface of C$^{18}$O $J=2-1$ from the \textsc{dali} model. 
\refchange{The region $z/r$ < 0.1 quickly becomes highly optically thick 
for C$^{18}$O $J=2-1$ with optical depths of $\tau_{\rm mm}$ = 10 near the midplane.}
For $z/r$ > 0.1, the C$^{18}$O abundance of the model would have been sensitive to changes 
in the gas temperature. Thus, where $z/r$ < 0.1, we can alter the disk
structure without affecting the intensity profiles of the CO isotopologues.

Five parameterized model scenarios are tested: \change{a disk with high midplane CO abundance, 
a disk with low midplane CO abundance, a cold disk, a shadowed cold disk, and a shadowed
cold disk including CO depletion. The aim is to initially test the CO abundance and the gas temperature
separately to investigate which parameter has a stronger influence on the
production of DCO$^+$. \newchange{These two parameters are intrinsically interlinked 
(i.e., more CO freeze-out will occur in lower temperature environments),
and the parameterized models allow us to explore these effects in isolation.}
In the parameterized models we also include an additional DCO$^+$ 
constant abundance region to act as a proxy for warm DCO$^+$ formation.}

\newchange{Fits to the radial intensity profile of the data are used to evaluate
the model parameters.} To obtain model intensity profiles,
synthetic DCO$^+$ image cubes are created using the 2D gas density, gas
temperature, and DCO$^+$ abundance structure from \textsc{dco+ chemnet}
as input to the \underline{LI}ne \underline{M}odeling \underline{E}ngine
\citep[\textsc{lime};][]{Brinch2010} radiative transfer code. \textsc{lime}
was run in LTE with 30000 grid points to create synthetic images of 
the DCO$^+$ $J=3-2$ transition. The images are continuum-subtracted and sampled
in the $uv$ plane using the \textsc{python}
\texttt{vis\_sample}\footnote{\texttt{vis\_sample} is publicly 
available at \url{https://github.com/AstroChem/vis\_sample} or in the 
Anaconda Cloud at \url{https://anaconda.org/rloomis/vis\_sample}} routine,
which reads the $uv$ coordinates directly from our observed ALMA measurement set
and creates synthetic visibilities of the model. The model visibilities
are imaged in \textsc{casa} using \textsc{clean} with natural weighting, and
an integrated intensity map was created over the same velocity range as 
the data (5.4 -- 8.8 km s$^{-1}$). Azimuthally-averaged elliptical annuli
projected to the disk inclination and position angle are used to extract the 
integrated intensity of the model with the same radial bins as the data.

\subsubsection{DCO$^+$ from deuterium fractionation}
\label{sec:mod_dcop_cold}

DCO$^+$ formation occurs as a result of deuterium fractionation, which is
an enhancement in the D/H ratio observed in certain deuterium-bearing 
molecules. Typically, deuterium fractionation occurs in colder environments
such as pre-stellar cores and below the surface layers of protoplanetary disks
because of the lower zero-point energies of the deuterated molecular ions \citep{Brown1986,Millar1989}.
\newchange{Deuterium fractionation in the low-temperature regime
occurs via the reaction \citep{Wootten1987}}

\begin{equation}
 \label{eq:h2dp_cold}
 \mathrm{HD \ + \ H_3^+ \ \longleftrightarrow \ H_2 \ + \  H_2D^+ \ + \ } \Delta E,
\end{equation}

\noindent where $\Delta E$ = 220 K \citep{Roberts2000,Gerlich2002,Albertsson2013}.
\newchange{This deuterium fractionation pathway is typically efficient at
temperatures below $\sim$30 K due to the energy barrier $\Delta E$
for the back-reaction. The DCO$^+$ molecule is then formed by the following reaction}

\begin{equation}
 \label{eq:dcop_cold}
 \mathrm{H_2D^+ \ + \ CO \ \longrightarrow \ H_2 \ + \  DCO^+ }.
\end{equation}

Gas-phase CO is needed to produce DCO$^+$, but CO will also rapidly combine with H$_3^+$, 
quenching the production of H$_2$D$^+$. Thus there is a balance where
CO must be sufficiently depleted for H$_2$D$^+$ to remain abundant
yet enough gas-phase CO must be present so that DCO$^+$ may form.
The simple chemical network from \citet{Murillo2015} is used
in this work to model DCO$^+$ produced via the cold 
deuterium fractionation pathway. The model shown in Figure~\ref{fig:mod_dens_temp_co}
suggests that \newchange{DCO$^+$ is expected throughout much
of the disk midplane, as temperatures are cold enough 
for CO freeze-out and the production of H$_2$D$^+$.}

\newchange{Deuterium fractionation in disks can also occur when HD and CH$_3^+$ 
combine to create H$_2$ and CH$_2$D$^+$ \citep{Millar1989}. 
DCO$^+$ is then formed via CH$_2$D$^+$ reactions directly or via one 
of its products, CH$_4$D$^+$. The energy barrier for the 
back-reaction of this deuterium fractionation pathway was recently 
revised \citep[$\Delta E$ = 654 K;][]{Roueff2013} and suggests that CH$_2$D$^+$,
and therefore DCO$^+$, could be formed efficiently at higher temperatures. 
In recent models of the TW Hya disk with a full deuterium chemical network,
\citet{Favre2015} find that T $\geq$ 71 K is sufficient to switch 
off the production of DCO$^+$ formed via CH$_2$D$^+$.}
The warmer CH$_2$D$^+$ fractionation route dominates over the cold H$_2$D$^+$ fractionation 
route for temperatures greater than $\sim$30 K because of the higher
energy barrier and the fact that H$_2$ will readily destroy H$_2$D$^+$ above $\sim$30 K. 

\begin{table*}[!htp]
 \caption{HD 169142 parameterized models.}
 \centering
 \label{tab:mod_colddisk}
 \resizebox{15cm}{!}{
 \begin{tabular}{lccccc}
 \hline \hline
 {\bf Model} & \multicolumn{2}{c}{\bf Disk Region} & \textbf{Temperature Profile} & \multicolumn{2}{c}{\bf Abundance Modifications} \\
 & $r$ & $z/r$ & $T$\tablefootmark{a} & CO factor\tablefootmark{b} & $X({\rm DCO^+_{warm}})$\tablefootmark{c} \\
 & [AU] &  & [K] & [$\times$CO] & \\
 \hline
 High CO & $\geq$83 & $<$0.1 & $T_{\rm gas}$  & 10 & 2.0$\times$10$^{-12}$\\
 Low CO & $\geq$83 & $<$0.1 & $T_{\rm gas}$  & 0.1 & 2.0$\times$10$^{-12}$\\
 Cold disk & $\geq$83 & $<$0.1 & $T_{\rm gas}-2$  & $-$ & 2.0$\times$10$^{-12}$\\
 Cold disk $-$ shadowed & 83--120 & $<$0.1 & $T_{\rm gas}-8$  & $-$ & 2.0$\times$10$^{-12}$\\
 & $>$120 & $<$0.1 & $T_{\rm gas}-2$ & $-$ & 2.0$\times$10$^{-12}$\\
 Cold disk $-$ shadowed & 83--120 & $<$0.1 & $T_{\rm gas}-8$  & 0.2 & 2.0$\times$10$^{-12}$\\
 and depleted & $>$120 & $<$0.1 & $T_{\rm gas}-2$ & $-$ & 2.0$\times$10$^{-12}$\\
 \hline
 \end{tabular}
 }
 \tablefoot{
 \tablefoottext{a}{$T_{\rm gas}$ is the two-dimensional gas temperature structure taken from the optimized \textsc{dali} model. (Figure~\ref{fig:mod_dens_temp_co})}
 \tablefoottext{b}{The CO factor is multiplied by the CO abundance structure taken from the optimized \textsc{dali} model. (Figure~\ref{fig:mod_dens_temp_co})}
 \tablefoottext{c}{The $X({\rm DCO^+_{warm}})$ component is included from 30--70 K.}
 }
\end{table*}

\subsubsection{CO abundance vs. gas temperature}
\label{sec:mod_colddisk_mods}

Table~\ref{tab:mod_colddisk} shows the parameters used for each model scenario. 
\change{The high CO and low CO models respectively increase and decrease the CO 
abundance in the region of interest ($r > 83$ AU and $z/r$ < 0.1) in order to 
determine how changes in the availability of gas-phase CO influence the DCO$^+$
emission. The cold disk model tests the influence of the gas temperature on the DCO$^+$
emission by moderately decreasing the temperature profile in the region of interest.
The shadowed cold disk model expands on the cold disk model by creating a \newchange{secondary} colder 
region just outside of the edge of the millimeter grains.}

\newchange{In this work we do not include a chemical network
describing the formation of DCO$^+$ via the warm deuterium fractionation 
pathway. \refchange{The complexity of hydrocarbon cation chemistry on which the warm
deuterium fractionation pathway depends introduces large uncertainties in the results
of even basic chemical networks. As a first-order approximation, we} instead 
adopt a region of constant DCO$^+$ abundance for 30 K$\ \leq T \leq$ 70 K}, 
\refchange{with the lower temperature limit based on the energy barrier of the 
back reaction for Equation~\ref{eq:h2dp_cold} and the upper temperature limit
based on the results of \citet{Favre2015}, where they find that formation of DCO$^+$
via the warm deuterium fractionation pathway is effectively switched off at 71 K.
In this way we introduce a proxy in the model} that is 
representative of DCO$^+$ production in the high-temperature regime, with
the reasonable expectation that the warmer pathway will contribute little to the emission
of the outer disk \citep{Oberg2015}, \refchange{which is the focus of this work}. Further investigation on the detailed
contribution of the warm deuterium fractionation pathway to the overall DCO$^+$ production,
with particular attention to the inner regions of the disk at radii $\lesssim$50 AU,
will be the focus of future work.

\newchange{The constant abundance component from 30--70 K 
is tuned so that the model radial profile matches 
the intensity of the observed DCO$^+$ emission from 40--70 AU, 
\refchange{where warm DCO$^+$ is the primary contributor before
peaking and turning over at 70--80 AU. This gives
$X({\rm DCO^+_{warm}})$ = $2.0 \times 10^{-12}$, which is 
consistent with the DCO$^+$ abundance found by \citet{Willacy2009} 
in models of a protoplanetary inner disk
at radii less than 30 AU, and consistent with the abundance 
between 30 -- 70 K in more recent work by \citet{Oberg2015} 
in their model of IM Lup. It is roughly one to two
orders of magnitude lower than the abundance found by 
\citet{Favre2015}, but their warm DCO$^+$ was confined to a 
thin layer spanning only 1 -- 2 AU at radii less than 60 AU.
In HD 169142 the 30 -- 70 K layer spans roughly 5 -- 15 AU, depending on
the radius, and results in a DCO$^+$ column density on the order of 10$^{12}$
cm$^{-2}$, which is consistent with \citet{Favre2015}.}
With this treatment, the warm deuterium fractionation pathway contributes $<$20\% 
to the DCO$^+$ radial intensity profile for $r > 83$ AU in all models considered. 
For $r < 83$ AU, the warm component is the primary contributor to the 
overall DCO$^+$, producing $>$80--95\% of the emission, depending on the model.
For $r \lesssim 30$ AU, no warm component of DCO$^+$ exists in the model because 
the midplane gas temperature reaches 70 K at $\sim$30 AU.
Because the warm component is set to the same \refchange{abundance value}
for all parameterized models, we consider the outer disk at $r > 83$ AU
for comparison between the data and the model. }

\begin{figure*}[!tp]
 \centering
 \includegraphics[width=\textwidth]{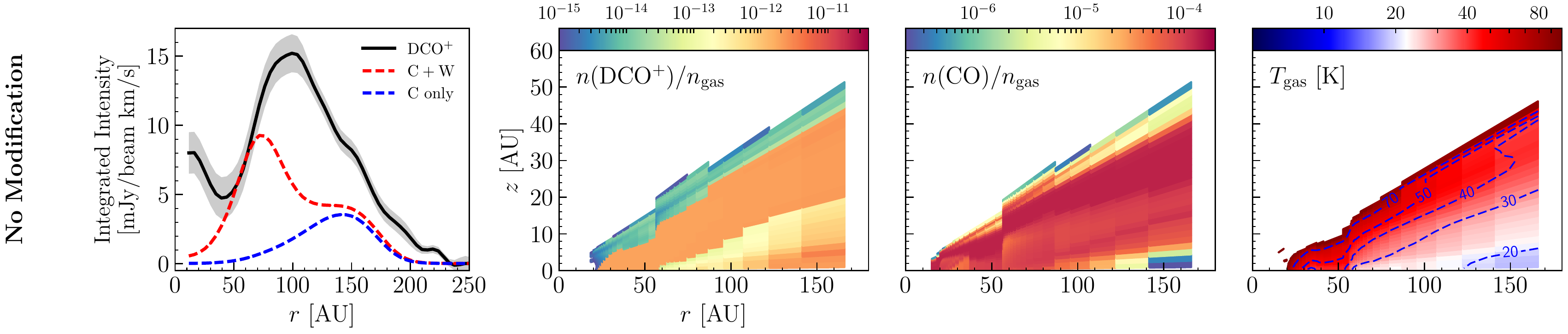} \\
 \vspace{0.2cm} 
 \includegraphics[width=\textwidth]{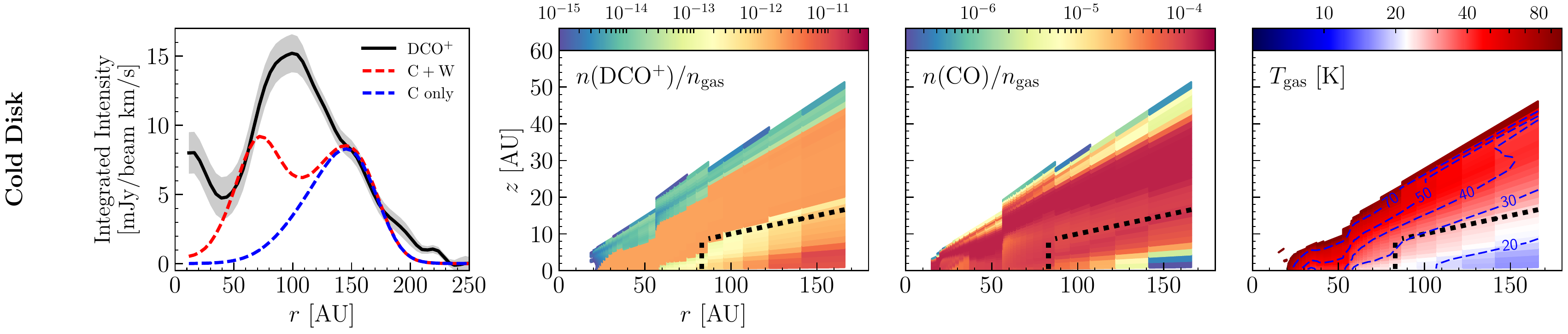} \\
 \vspace{0.2cm}
 \includegraphics[width=\textwidth]{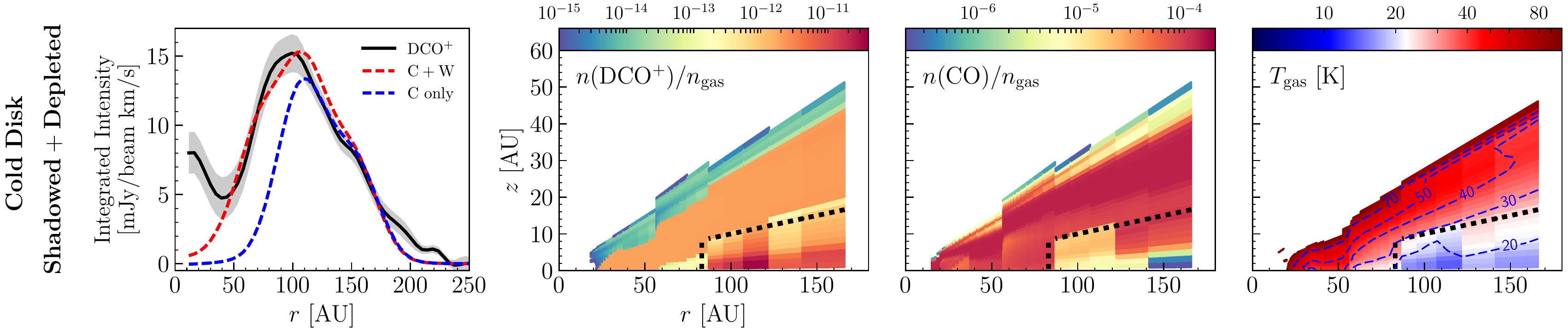} \\
 \caption{Parameterized modifications to the optimized \textsc{dali} model
 described in \newrefchange{Appendix~\ref{app:A}}.
 The right and right-middle plots show the gas temperature 
 and CO abundance, respectively. 
 The black dashed box shows the modified region. 
 The left-middle plot shows the DCO$^+$ 
 abundance map calculated with \textsc{dco+ chemnet} with the 30--70 K 
 constant abundance warm component included. 
 The left plot shows the DCO$^+$ radial intensity profiles of the 
 data (solid black with shaded gray 1$\sigma$ errors) and of the model
 with the cold component only (C only; dashed blue) and with warm 
 component included (C + W; dashed red).
 ({\it Top}) The optimized \textsc{dali} model with no modifications made to the
 gas temperature or CO profile.
 ({\it Middle }) Cold disk model. The gas temperature is decreased by 2 K
 in the black dashed region.
 ({\it Bottom}) Shadowed and depleted cold disk model. An extension of the cold 
 disk model to simulate a secondary colder region beyond the millimeter dust edge.
 The gas temperature is decreased by 8 K with CO depleted by a factor of five 
 from $r = 83 - 120$ AU. The gas temperature is decreased by 2 K for $r > 120$ AU.}
 \label{fig:dcop_chemnet}
\end{figure*}

\newchange{Figure~\ref{fig:dcop_chemnet} 
shows the gas temperature map and $^{12}$CO abundance map used as input, 
the DCO$^+$ abundance map calculated by \textsc{dco+ chemnet}
\refchange{and including the constant abundance warm component}, 
and the DCO$^+$ radial intensity profiles derived from the LIME synthetic images
with only the cold deuterium fractionation pathway active (C only) and with 
the constant abundance warm component included (C + W).
The no modifications model in the first row of Figure~\ref{fig:dcop_chemnet} 
uses the input from our optimized \textsc{dali} model with no alterations to the
midplane disk structure. It is already clear that the 
DCO$^+$ emission is under-produced by the \textsc{dco+ chemnet} 
code for the disk structure outlined in \newrefchange{Appendix~\ref{app:A}.}}

In the low CO abundance case, \newchange{an order of magnitude decrease in midplane gas-phase
CO abundance results in a factor of two to three decrease in} DCO$^+$ emission because there is no 
longer enough gas-phase CO available to efficiently form DCO$^+$. 
\newchange{The high CO abundance model with an one order of magnitude
increase in gas-phase CO abundance results in only a factor of two increase in 
DCO$^+$ emission, highlighting the non-linear nature of the chemistry. 
CO abundances in this model become much higher than the 
canonical value of $\sim$10$^{-4}$ with respect to H$_2$ in the midplane of disk
where CO depletion is expected; hence, this model is physically unrealistic. 
Order of magnitude} variations to the CO abundance do not have a significant influence
on the formation of DCO$^+$ near the outer disk midplane for the 
physical structure given by the optimized \textsc{dali} model. 

\newchange{The second and third rows of Figure~\ref{fig:dcop_chemnet}
illustrate increasingly colder midplane scenarios.}
In the cold disk model case, \newchange{a decrease of 2 K in the gas temperature}
provides an improved fit to the DCO$^+$ radial profile but still fails
to fully recover much of the DCO$^+$ between 80--140 AU. It matches
the observed intensity profile well beyond 150 AU.
The shadowed cold disk model expands on the cold disk model to 
invoke a secondary cold region from $r = 83-120$ AU with $T = T_{\rm gas} - 8$ K. 
\newchange{For the secondary cold region, the outer boundary and temperature drop parameters 
are explored from 100--140 AU in steps of 10 AU and from $T_{\rm gas} - 3$ K 
to $T_{\rm gas} - 10$ K in steps of 1 K, respectively. The outer boundary at 120 AU and the temperature drop
of $T_{\rm gas} - 8$ K was found to provide the best improvement on the fit to the DCO$^+$ profile.}
More DCO$^+$ is produced in the shadowed cold disk model between 80--150 AU
than in the cold disk model case, but it is still not enough 
to capture the 100 AU peak. The final model invokes CO depletion 
\newchange{in the secondary cold region of the} shadowed 
cold disk model, which would be expected for significantly colder disk regions. 
\newchange{The CO abundance is reduced by a factor of five between $r = 83-120$ AU.
CO depletion factors of two, five, and ten are tested, with a factor of five 
resulting in the right amount of DCO$^+$ emission.
In this scenario,} the DCO$^+$ radial intensity profile, including the 100 AU peak, is reproduced well. 
% The shadowed cold disk with CO depletion just beyond 
% the millimeter grains provides the best scenario 
% in this work to reproduce the observed DCO$^+$ in the disk around HD 169142.

% The underlying assumption in the parameterized CO abundance models
% was that there is a mechanism to account for the alterations in abundance. 
% For example, the changes could be indicative of variations in the efficiency of CO
% freeze-out or in the efficiency of CO desorption from ices in the area of the 
% disk where we are blind to the midplane conditions based only on the 
% CO isotopologue observations.
% Similarly, in the cold disk models there is an assumed mechanism to account for 
% sufficient cooling of the disk near the midplane. The number and distribution of small grains is
% likely not well-constrained near the midplane of the outer disk, which could
% lower the dust and gas temperature by a few Kelvin. A dramatic decrease in the 
% gas temperature just beyond the outer dust ring could be caused
% by shadowing of the outer disk at the millimeter dust edge. 

The efficiency of the cold deuterium fractionation pathway, 
and thus the production of DCO$^{+}$, is also affected by cosmic ray ionization, $\zeta_{\rm cr}$,
and the ortho-to-para (o/p) ratio of H$_2$. Ionization of H$_2$ in the disk by 
cosmic rays will affect the number of H$_3^+$ ions and the number of free electrons. 
The \textsc{dco+ chemnet} code was rerun for the model with no modifications to test changes
to the cosmic ray ionization rate, initially set to $\zeta_{\rm cr} = 1.26 \times 10^{-17}$ s$^{-1}$. 
\refchange{There is evidence that the local ISM in the HD 169142 region may reach 
values of $\zeta_{\rm cr} = 1-5 \times 10^{-16}$ s$^{-1}$ \citep{Indriolo2007,Neufeld2017}.
An increase in $\zeta_{\rm cr}$ of one order of magnitude results in DCO$^{+}$ emission 
comparable to the cold disk model, meaning that the amount of DCO$^{+}$ observed 
at $r > 120$ AU may be a consequence of a higher local cosmic ray ionization rate or 
may be due to a moderate decrease in the gas temperature of the outer disk.}
% However, there is currently no evidence that HD 169142 resides in a region of higher cosmic 
% ray flux than found in the canonical local ISM.
Ionization of the outer disk by UV radiation may influence the cold deuterium fractionation
pathway and free electron population differently, but the effect is not modeled here.
Such ionization would influence primarily the disk upper layers, as UV photons will not reach 
the disk midplane with a sufficient flux to outpace ionization due to cosmic rays.

The o/p ratio of H$_2$ influences the survival of H$_2$D$^+$ in cold regions of the disk
because the back-reaction in Equation~\ref{eq:h2dp_cold} has a lower energy
barrier for o-H$_2$ ($\Delta E$ = 61 K) than for p-H$_2$ \citep[$\Delta E$ = 232 K;][]{Walmsley2004}.
We apply a thermal treatment of the o/p ratio of H$_2$ in LTE following \citet{Murillo2015} with 
${\rm o/p} = 9 \times {\rm exp}(-170/T)$, where $T$ is the gas temperature, \refchange{and again
rerun \textsc{dco+ chemnet} code for the no modification model including 
the H$_2$ o/p ratio, which was not considered in the original models.} In this case,
too much DCO$^{+}$ emission is produced in the outer disk for $r > 150$ AU, but 
DCO$^{+}$ is under-produced for $r < 150$ AU. However, this requires an efficient 
equilibration of the H$_2$ spin temperature with the gas temperature, rather than
the 3:1 obtained from grain surface formation of H$_2$.
The degree of cosmic ray ionization
and the precise distribution of the o/p ratio of H$_2$ will affect the exact values
for the temperature drop and CO depletion necessary to obtain a fit to the DCO$^{+}$
radial intensity profile, but our overall conclusions remain unchanged.

Based on the best-fit shadowed \newchange{and depleted} cold disk model, a 
significantly colder outer disk midplane \newchange{with increased CO freeze-out}
is the most likely scenario for DCO$^{+}$ production beyond the 
millimeter grains. \newchange{The DCO$^{+}$ emission at $r > 120$ AU could 
be reproduced with a small drop in the gas temperature or an increase in the 
\refchange{cosmic ray} ionization rate.}
The modeling efforts presented here show
that DCO$^{+}$ emission can reveal \newrefchange{structure in}
low-temperature regions in the midplane of disks that are not apparent 
in \newrefchange{the CO isotopologues}.

\section{Discussion}
\label{sec:disc}

The interpretation of DCO$^+$ emission regardings its chemical origins
and location within the disk
is complicated by the multiple deuterium fractionation pathways
available. The two main pathways (cold, via H$_2$D$^+$
fractionation and warm, via CH$_2$D$^+$ fractionation) are efficient
over different temperature ranges, therefore it is useful to consider
DCO$^+$ emission from distinct regions of the disk where the 
conditions are expected to be more favorable for one fractionation
pathway over the other. The inner disk provides warmer temperatures
that can switch off the H$_2$D$^+$ pathway, while the outer disk hosts
a cold midplane that allows the H$_2$D$^+$ pathway to operate efficiently.

\subsection{Inner disk DCO$^+$}
\label{sec:disc_inner_dcop}

In models of TW Hya including a full deuterium chemical network, 
\citet{Favre2015} found that DCO$^+$ observed in the inner tens of AU 
is not primarily formed by the 
cold deuterium fractionation pathway via H$_2$D$^+$ because
of the warm temperatures of the inner disk. The physical structure 
in Figure~\ref{fig:mod_dens_temp_co} shows that the disk around HD 169142 is
far too warm in the inner 50 AU for the H$_2$D$^+$ fractionation
pathway to be the main contributor. Instead, DCO$^+$ in this region is likely formed by the 
warmer CH$_2$D$^+$ fractionation pathway. 

The disk temperature at the midplane is greater than 70 K at $r<30$ AU,
therefore even the warm component is switched off. In this disk, DCO$^+$ formed
via the CH$_2$D$^+$ fractionation pathway may continue to be active
at temperatures greater than 70 K. \change{Alternatively, we may be missing a cool inner
component, such as inner disk dust rings that \newchange{could keep temperatures low
enough for the warm deuteration fractionation pathway to remain active.} 
\citet{Andrews2016} observed optically thick millimeter dust rings on the
order of a few AU in the disk around TW Hya. Recent work by \citet{Ligi2018}
presented a tentative detection of another dust ring in the HD 169142 disk
located at $\sim$0.1$\arcsec$ [12 AU at 117 pc] using VLT/SPHERE radial differential 
imaging with the IRDIS and IRF instruments.}

The molecule DCN provides another avenue to probe the \newchange{warm component
of DCO$^+$ emission} as it is also formed via warm deuterium fractionation in disks \citep{Millar1989}. 
\newchange{Co-spatial peaks in DCO$^+$ and DCN would indicate that the 
warm deuterium fractionation pathway is a strong contributor to the production of DCO$^+$. 
Recent observations of DCN and DCO$^+$ in several T Tauri and Herbig Ae/Be sources
show them peaking in different regions of the disk, but with some DCO$^+$ present 
where the DCN peaks, indicating that the warm deuterium fractionation pathway 
contributed partially to the DCO$^+$ emission \citep{Qi2008,Oberg2012,Huang2017,Salinas2017}.}

\subsection{Outer disk DCO$^+$}
\label{sec:disc_outer_dcop}

In order to recover the observed DCO$^+$ radial \change{intensity} profile,
it was necessary to modify the structure of the optimized
\textsc{dali} model from Figure~\ref{fig:mod_dens_temp_co} to include a 
much colder, CO depleted region beyond the edge of the millimeter grains.
The cause of the decrease in temperature 
could be due to the number and location of micron-sized grains
in the outer disk, which have been observed out to at least 200 AU in
scattered light imaging \citep{Quanz2013,Momose2015,Monnier2017,Pohl2017}.
\newchange{The exact, local distribution of the micron-sized grains in the HD 169142
disk might be different than the distribution approximated in the optimized 
\textsc{dali} model, leading to lower gas temperatures in the outer disk.}

The fact that the peak in the DCO$^+$ radial intensity profile occurs
at 100 AU is already evidence that the colder H$_2$D$^+$ deuterium 
fractionation pathway is responsible for the majority of DCO$^+$ production
in this disk. As mentioned in Section~\ref{sec:disc_inner_dcop}, the 
warm component is not significant \newchange{beyond the edge of the millimeter grains ($r = 83$ AU)}, with DCO$^+$
formed via the cold network in \textsc{dco+ chemnet} contributing
$>$80\% to the radial \change{intensity} profile. In this context,
it is \newchange{probable} that the 100 AU peak observed in the 
DCO$^+$ profile is tracing the CO snow line as CO begins to freeze out
onto dust grains, but is still sufficiently abundant in the gas-phase
to allow formation of DCO$^+$.  
\citet{Macias2017} previously compared the C$^{18}$O and DCO$^+$ radial intensity
profiles and similarly concluded that a co-incident slope change in the C$^{18}$O and 
peak in the DCO$^+$ is evidence that DCO$^+$ is tracing the CO snow line, which
they locate at $\sim$100 AU. The modeling done in this work shows that
invoking a cold, CO depleted region \newchange{in the outer disk midplane from $\sim$80--120 AU}
is necessary to recover the DCO$^+$ radial intensity profile in 
the parameterized models, \newchange{indicating that the bulk of 
the DCO$^+$ does indeed trace regions of CO freeze-out in the HD 169142 disk.}

\newchange{There is tentative evidence for an outer DCO$^+$ ring, seen as
a bump in the DCO$^+$ radial profile at the edge of the gaseous disk around 200 AU.
Similar DCO$^+$ structure has been seen in other disks, e.g., IM Lup, LkCa 15,
and HD 163296 \citep{Oberg2015,Huang2017,Salinas2017,Flaherty2017}. An outer ring could
indicate that some CO ice is returned to the gas phase in the outer disk
via thermal desorption caused by a temperature inversion or via
photodesorption by UV radiation \citep{Huang2016,Cleeves2016a,Cleeves2016b,Facchini2017}. 
The observed bump is weaker than the 
outer DCO$^+$ ring in other disks, suggesting that the HD 169142 
disk remains cold out to large radii or that the degree of 
CO freeze-out in this disk may not be as extreme as in other sources.}

\newchange{Follow-up observations of N$_2$H$^+$ in this disk could provide further insight
into the relationship between DCO$^+$ and CO freeze-out.
Formation of N$_2$H$^+$ requires low temperatures and significant CO depletion, and therefore
is expected to be abundant only in the cold midplane where there is a high 
degree of CO frozen out onto the surface of dust grains \citep{Walsh2012,Qi2015,vantHoff2017}.
Co-spatial emission of N$_2$H$^+$ and DCO$^+$ in the outer disk would place strong constraints
on the extent to which DCO$^+$ is directly tracing CO freeze-out.}

\section{Conclusions}
\label{sec:concl}

In this paper we present $\sim$0.3$\arcsec$ resolution ALMA observations of DCO$^+$ $J = 3-2$
in the protoplanetary disk around HD 169142. \change{We update the
fiducial \textsc{dali} model from \citet{Fedele2017a} to include small dust
grains throughout the disk and employ
a simple deuterium chemical network to investigate the production
of DCO$^+$ formed by the cold deuterium 
fractionation pathway (via H$_2$D$^+$). The CO abundance and gas temperature structure of the 
\textsc{dali} disk model is adapted in a parameterized way 
using the \textsc{dco+ chemnet} code to recover the observed 
DCO$^+$ radial intensity profile. The warm deuterium fractionation pathway 
(via CH$_2$D$^+$) is approximated with a constant abundance between
30--70 K in the parameterized models.} The conclusions of this work are the following:

\begin{itemize}
  \item \change{DCO$^+$ has a broad, ring-like morphology over radii $\sim$50 -- 230 AU in the disk around HD 169142,
  \newchange{with most of the emission located outside of the millimeter continuum edge. 
  There is an inner component to the DCO$^+$ radial profile with emission returning in 
  the central $\sim$30 AU. The DCO$^+$ radial intensity profile peaks at $\sim$100 AU, 
  with a tentative secondary bump at $\sim$200 AU.}}
%   \item The \textsc{dali} model with a simple chemical network to form DCO$^+$ 
%   from cold deuterium fractionation
%   under-produces the observed emission by \change{one order of magnitude, indicating
%   that the disk midplane environment is not properly described by the \textsc{dali} model.}
%   \item \change{PAHs as a destruction mechanism for H$_2$D$^+$ and DCO$^+$ should be considered
%   in simple chemical modeling of DCO$^+$ formation. If the PAH abundance structure is not 
%   well-constrained, the production of DCO$^+$ can vary by an order of magnitude or more 
%   depending on the adopted values.}
  \item \newchange{Parameterized modeling of the HD 169142 disk shows that 
  lowering the gas temperature in optically thick regions near the midplane of the 
  outer disk by a several K has a significant effect on the DCO$^+$ profile. Order of magnitude
  changes to the CO abundance in the same region cause the DCO$^+$ radial 
  profile to increase or decrease only by a factor of 2--3, still under-producing the amount of observed DCO$^+$.
  It was necessary to invoke both effects to successfully reproduce the full radial intensity profile.}
  \item The best-fit parameterized model has a shadowed, cold region with CO depletion 
  \newchange{near the disk midplane just beyond the edge of the millimeter grains}. For $z/r$ < 0.1 and 
  $r > 83$ AU, the model recovers the radial intensity profile of DCO$^+$
  with a $T = T_{\rm gas} - 8$ K region with a factor of five CO depletion from 
  $r = 83 - 120$ AU, and a $T = T_{\rm gas} - 2$ K region for $r > 120$ AU. 
  \newchange{The exact values for the drop in $T_{\rm gas}$ and CO abundance will
  also depend on the ionization degree and the ortho-to-para ratio of H$_2$.} The fact that the 
  added shadowed region is needed to recover the 100 AU radial intensity peak highlights the 
  sensitivity of DCO$^+$ to small changes in the gas temperature and CO abundance structure.
  \item The best-fit model suggests that the contribution to the overall
  DCO$^+$ emission from the cold deuterium fractionation pathway via H$_2$D$^+$
  is $>$85\% at $r > 83$ AU, \change{while the 
  contribution from the warm deuterium fractionation pathway via CH$_2$D$^+$ is $>$80\% 
  at radii less than $r < 83$ AU using a constant abundance of
  $X({\rm DCO^+_{warm}}) = 2.0 \times 10^{-12}$ \newchange{from 30--70 K}. The warm component
  does not recover the return of DCO$^+$ emission within $\sim$30 AU for
  the current disk structure.}
%   \item The optimized \textsc{dali} model \newchange{used for the physical structure
%   of the disk in this work} has a midplane region beyond the millimeter
%   dust edge that is too warm by several Kelvin. A colder region could be realized 
%   by adjusting the distribution of the micron-sized grains and/or by
%   \newchange{shadowing of the outer disk at the location of the millimeter dust edge.}
  \item DCO$^+$ is an optically thin molecular tracer that acts 
  as a filter to detect disk substructure that is not observable in more abundant,
  and therefore more easily detectable, molecular tracers such as $^{12}$CO, 
  $^{13}$CO, and C$^{18}$O. \newchange{In this work, DCO$^+$ observations
  reveal the low-temperature ($\lesssim$25 K) midplane structure in the disk around HD 169142.}
  
\end{itemize}

This work shows that DCO$^+$ can be used as a valuable tracer of protoplanetary 
disk midplane conditions with simple models and chemical networks.
To fully characterize the complex chemical origins of DCO$^+$ in the disk
around HD 169142, full dust evolution models \citep[e.g.,][]{Facchini2017}
and an expanded deuterium chemical network such as the one used by \citet{Favre2015}
would be required. Further constraints could be placed on the origins of DCO$^+$
with future ALMA observations of additional chemical tracers, such as DCN and 
N$_2$D$^+$, that would correlate with DCO$^+$ formed via the warm and cold
deuterium fractionation pathways, respectively.

\begin{acknowledgements}
      \newchange{
      The authors thank the anonymous referee for useful comments
      that helped to improve the paper.
      M.T.C. thanks A. Pohl for providing the VLT/SPHERE $J$-band
      polarized intensity data.
      M.T.C. and M.R.H. acknowledge support from the
      Netherlands Organisation for Scientific Research (NWO)
      grant 614.001.352. C.W. acknowledges the NWO
      (grant 639.041.335) and the University of Leeds for financial support.
      D.F. and C.F. acknowledge support from the Italian Ministry of Education, 
      Universities and Research, project SIR (RBSI14ZRH).}
      This paper makes use of the following ALMA data: ADS/JAO.ALMA\#
      2013.1.00592.S. 
      ALMA is a partnership of ESO (representing its member states), 
      NSF (USA) and NINS (Japan), together with NRC (Canada), NSC and 
      ASIAA (Taiwan), and KASI (Republic of Korea), in cooperation with 
      the Republic of Chile. The Joint ALMA Observatory is operated by 
      ESO, AUI/NRAO and NAOJ. A.M. acknowledges an ESO Fellowship.
\end{acknowledgements}

%-------------------------------------------------------------------

\bibliographystyle{aa}
\bibliography{carney_phd_paper2}

%-------------------------------------------------------------------
\clearpage
\begin{appendix}

  \newpage
  \section{Model details}
  \label{app:A}

  Modifications to the fiducial \textsc{dali} model
  from \citet{Fedele2017a} were necessary to more accurately incorporate
  the (sub)micron-sized grain population into the HD 169142 disk model and to obtain a better
  description of the dust and gas temperature of the outer disk beyond the 
  millimeter continuum edge. Here the modifications are discussed in more detail.
  
  The \citet{Fedele2017a} model fit the 1.3 mm dust emission, and
  dust in their models exists only within dust rings from $r$ = 20--35 AU and from $r$ = 56--83 AU.
  There is no dust of any kind in their models outside of these radii.
  \newchange{Observations by \citet{Quanz2013}, \citet{Monnier2017}, and \citet{Pohl2017} have revealed
  the presence of a micron-sized dust population in the millimeter ring gap 
  and in the outer disk beyond the millimeter-sized dust population.}
  \citet{Monnier2017} discussed an outer disk in the context of micron-sized
  grains observed in the $J$-band with the Gemini Planet Imager.
  The authors detected a dip in polarized light at $\sim$55 AU which they model
  as a 40--70 AU gap, and they observed a flared outer ring that peaks at 
  $\sim$75 AU. The authors adopted a distance of 145 pc for HD 169142. Their radii 
  scaled to the new Gaia distance of 117 pc gives a 32--56 AU gap 
  and a 60 AU outer ring peak in the $J$-band, consistent
  with the values for the millimeter gap and outer ring observed by \citet{Fedele2017a}.

  \newchange{The original dust structure of the \citet{Fedele2017a} model was}

  \begin{equation}
  \label{eq:surfdens_dust}
    n_{\rm dust}=\begin{cases}
      0 & \ {\rm for} \ r < R_{\rm dust \ in} \\
      \delta_{\rm dust} \times n_{\rm dust} & \ {\rm for} \ R_{\rm dust \ in} < r < R_{\rm gap \ in} \\
      0 &  \ {\rm for} \ R_{\rm gap \ in} < r < R_{\rm gap \ out} \\
      n_{\rm dust} & \ {\rm for} \ R_{\rm gap \ out} < r < R_{\rm dust \ out} \\
      0 & \ {\rm for} \ r > R_{\rm dust \ out}. \\
    \end{cases}
  \end{equation}

  \begin{table}[!hbp]
    \caption{HD 169142 \textsc{dali} model parameters.}
    \centering
    \label{tab:mod_params}
    \resizebox{!}{4.5cm}{
    \begin{tabular}{lll}
    \hline \hline
    Parameter & Value & Ref. \\
    \hline
    $M_{*}$ & 1.65 $M_{\odot}$ & 1 \\
    $T_{\rm eff}$ & 8400 K & 2 \\
    $L_{*}$ & 10 $L_{\odot}$ & 3 \\
    $d$ & 117 pc & 3 \\
    $i$ & 13$^\circ$ & 3,4 \\
    ${\rm P.A.}$ & 5$^\circ$ & 3,4 \\
    $\chi$ & 0.2 & 3 \\
    $f_{\rm large}$ & 0.85 & 3 \\
    $\psi$ & 0 & 3 \\
    $\gamma$ & 1 & 3 \\
    $h_{\rm c}$ & 0.07 & 3 \\
    $R_{\rm c}$ & 100 AU & 3 \\
    $\Delta_{\rm gd}$ & 80 & 3 \\
    $\Sigma_{\rm c}$ & 6.5 g cm$^{-2}$ & 3 \\
    $R_{\rm gas \ in}$ & 13 AU & 3 \\
    $R_{\rm dust \ in}$ & 20 AU & 3 \\
    $R_{\rm gap \ in}$ & 35 AU & 3 \\
    $R_{\rm gap \ out}$ & 56 AU & 3 \\
    $R_{\rm dust \ out}$ & 83 AU & 3 \\
    $R_{\rm gas \ out}$ & 180 AU & 3 \\
    $\delta_{\rm dust}$ & 0.27 & 3 \\
    $\delta_{\rm gas \ cavity}$ & 0.025 & 3 \\
    $\delta_{\rm gas \ gap}$ & 0.025 & 3 \\
    \hline
    \end{tabular}
  }
  \tablefoot{{\it References.} 1: \citet{Blondel2006}, 2: \citet{Dunkin1997},
  3: \citet{Fedele2017a}, 4: \citet{Raman2006}.
  }
  \end{table}

  \noindent However, the gap is not empty and the micron-sized grains are detected
  throughout the outer disk, thus their fiducial \textsc{dali} model was neglecting the presence of
  micron-sized grains that would insulate the outer disk midplane
  and thus lower dust and gas temperatures.

%   \noindent where $R_{\rm dust \ in}$ = 20 AU, the depletion factor $\delta_{\rm dust}$ = 0.27,
%   $R_{\rm gap \ in}$ = 35 AU, $R_{\rm gap \ out}$ = 56 AU, and $R_{\rm dust \ out}$ = 83 AU.
  
  We modified the dust structure of the \citet{Fedele2017a} fiducial model to expand the small grain 
  ($n_{\rm dust,small}$; 0.005 -- 1 $\mu$m) dust population to be present within the gap between the dust rings
  ($r = $ 35--56 AU) and at radii beyond the dust rings ($r > 83$ AU).
  In order to fit the SED, we included a thin inner region of hot dust at $r$ = 0.2 AU, 
  as observed by \citet{Wagner2015}. The optimized dust structure of the model is
  
  \begin{figure}[!tp]
    \centering
      \includegraphics[width=0.4\textwidth]{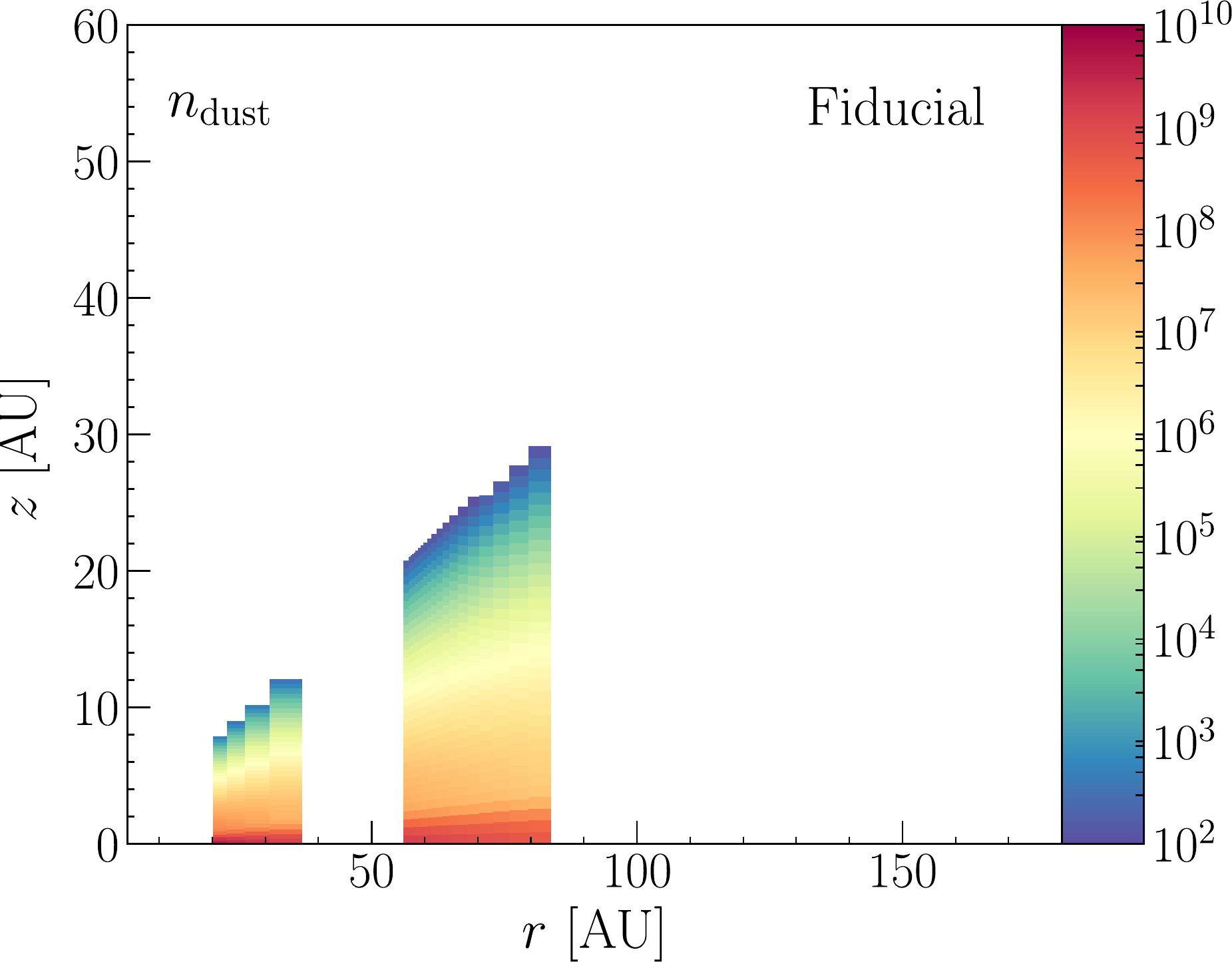} \\
      \vspace{0.5cm}
      \includegraphics[width=0.4\textwidth]{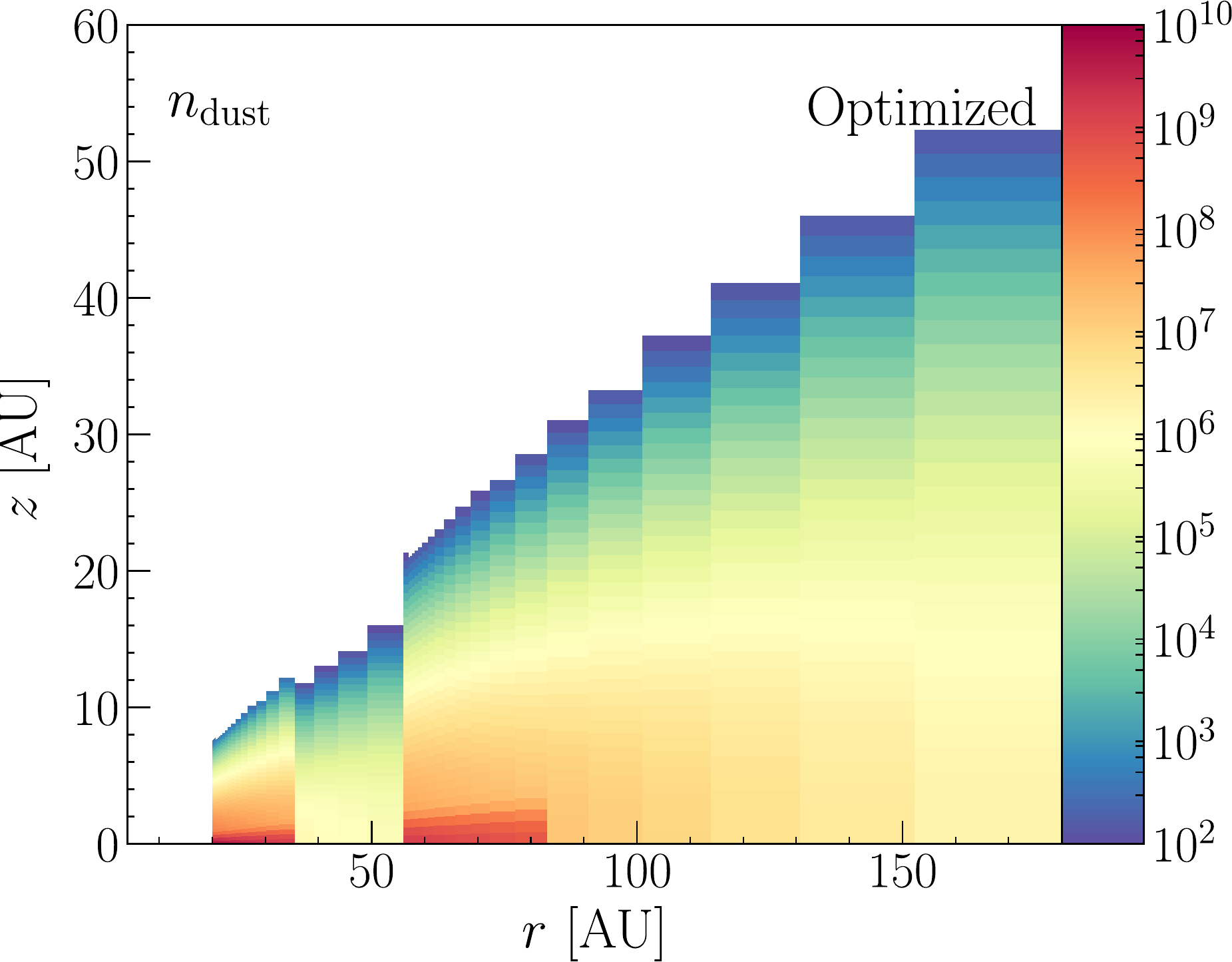} 
      \caption{({\it Top}) Dust structure of the fiducial \textsc{dali} model 
      from \citet{Fedele2017a} with small (0.005 -- 1 $\mu$m) and large 
      (0.005 -- 1000 $\mu$m) grains present only in the millimeter rings.
      ({\it Bottom}) Dust structure of the optimized \textsc{dali} model with 
      small grains in the millimeter gap and in the outer disk. The hot inner
      dust ring is not visible due to the extremely small scale height at $r = 0.2$ AU.
      }
  \label{fig:mod_dust_struc}
  \end{figure}

  \begin{equation}
  \label{eq:fid_surfdens_dust_new}
    n_{\rm dust}=\begin{cases}
    10^{-5} \times n_{\rm dust,small} & \ {\rm for} \ 0.2 < r < 0.4 \ {\rm AU} \\
    0 \ {\rm otherwise} & \ {\rm for} \ r < R_{\rm dust \ in} \\
    \delta_{\rm dust} \times n_{\rm dust} & \ {\rm for} \ R_{\rm dust \ in} < r < R_{\rm gap \ in} \\
    10^{-2} \times n_{\rm dust,small} &  \ {\rm for} \ R_{\rm gap \ in} < r < R_{\rm gap \ out} \\
    n_{\rm dust} & \ {\rm for} \ R_{\rm gap \ out} < r < R_{\rm dust \ out} \\
    n_{\rm dust,small} & \ {\rm for} \ r > R_{\rm dust \ out}. \\
    \end{cases}
  \end{equation}

  \begin{figure*}[!t]
    \begin{center}
      \includegraphics[width=0.28\textwidth,height=0.25\textwidth]{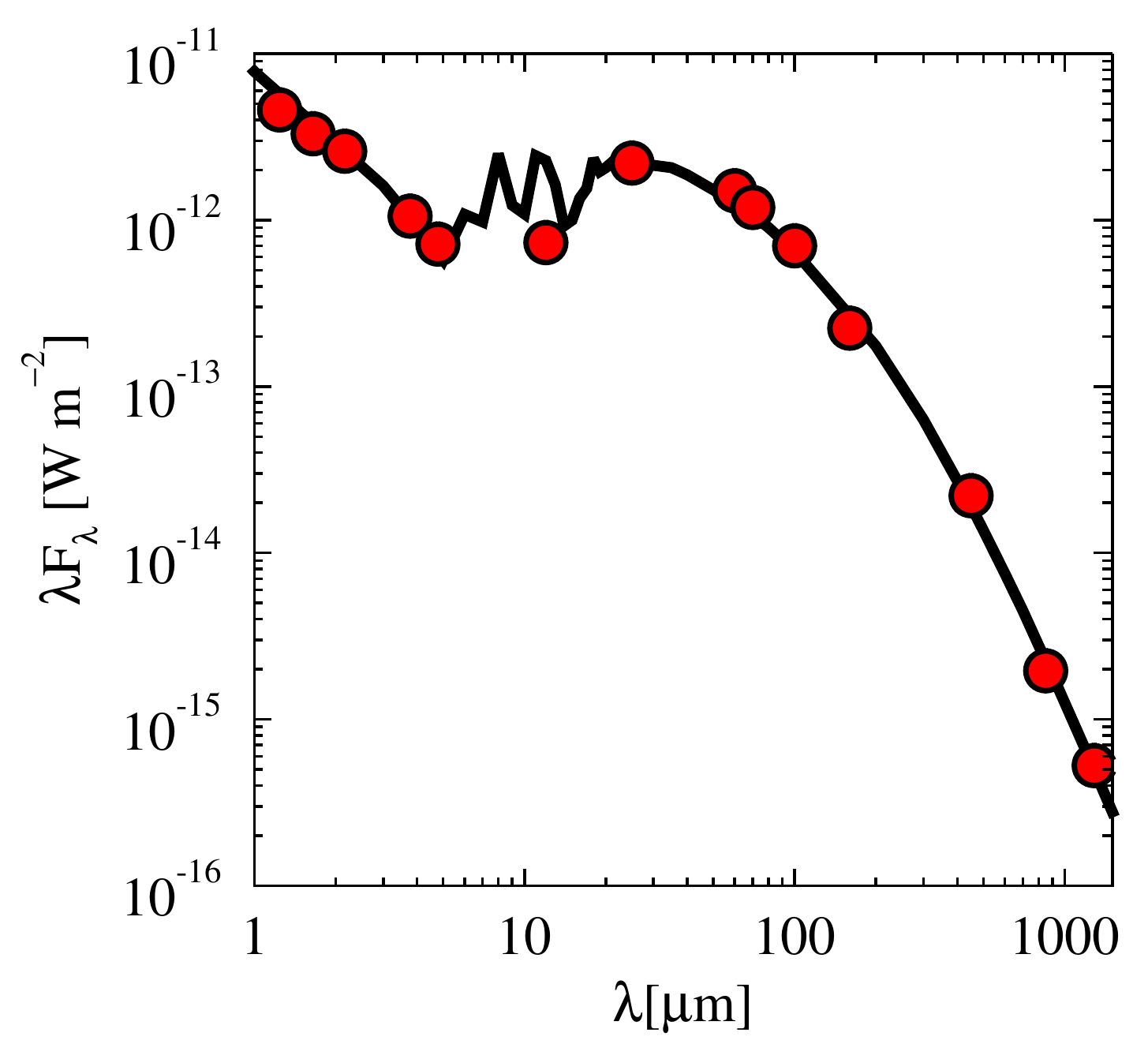}
      \hspace{0.3cm}
      \includegraphics[width=0.3\textwidth,height=0.25\textwidth]{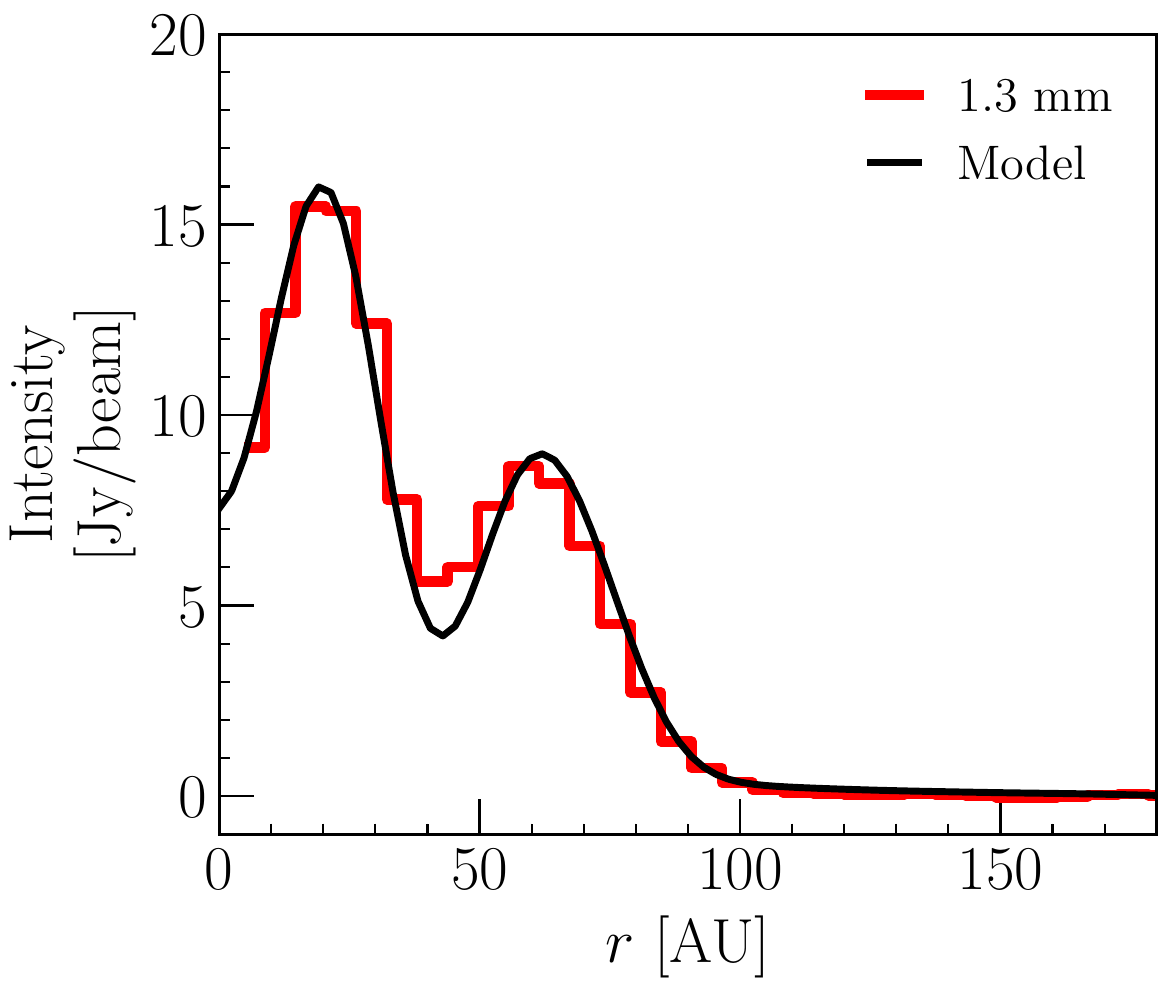} 
      \hspace{0.3cm}
      \includegraphics[width=0.3\textwidth,height=0.248\textwidth]{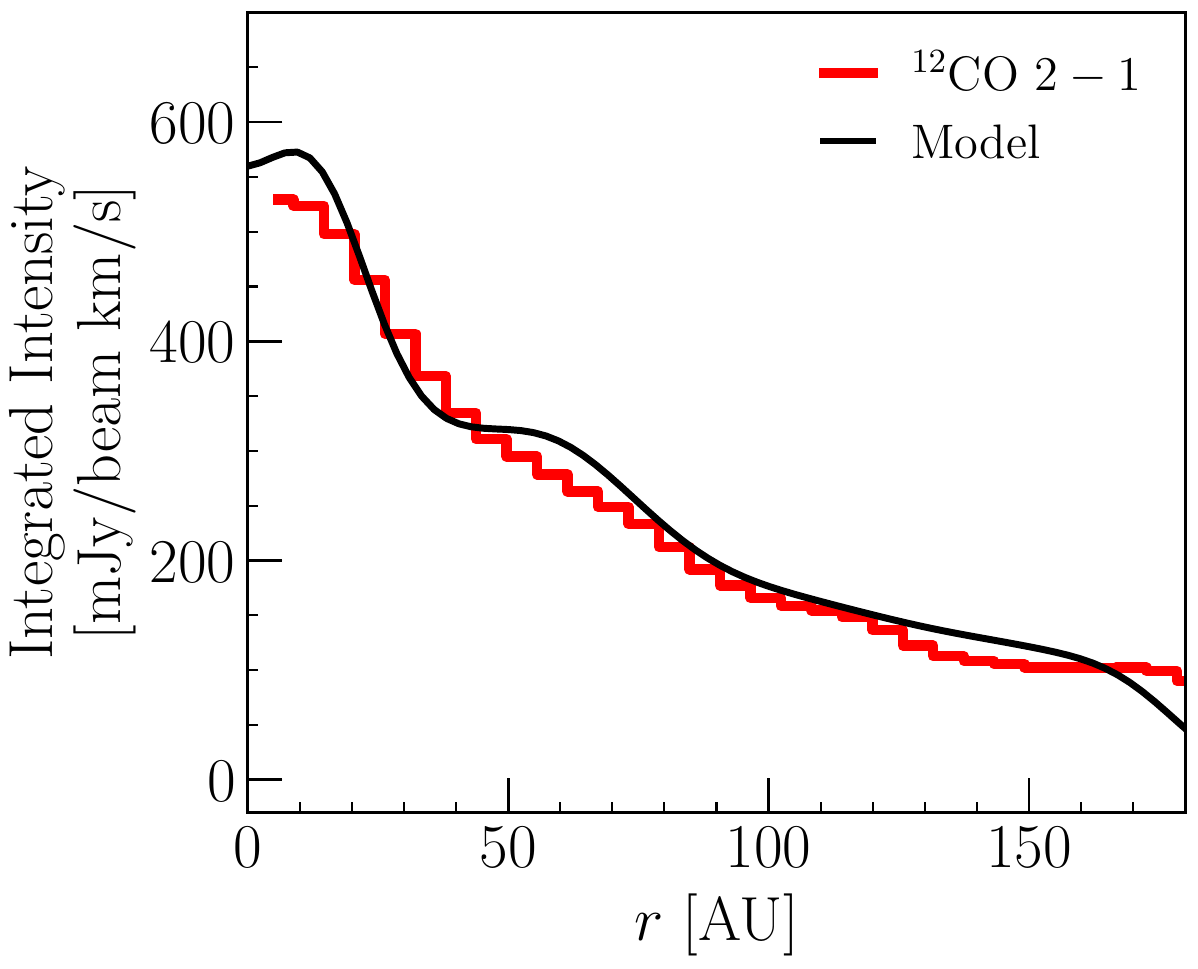} \\
      \vspace{1cm}
      \hspace{5.6cm}
      \includegraphics[width=0.3\textwidth,height=0.25\textwidth]{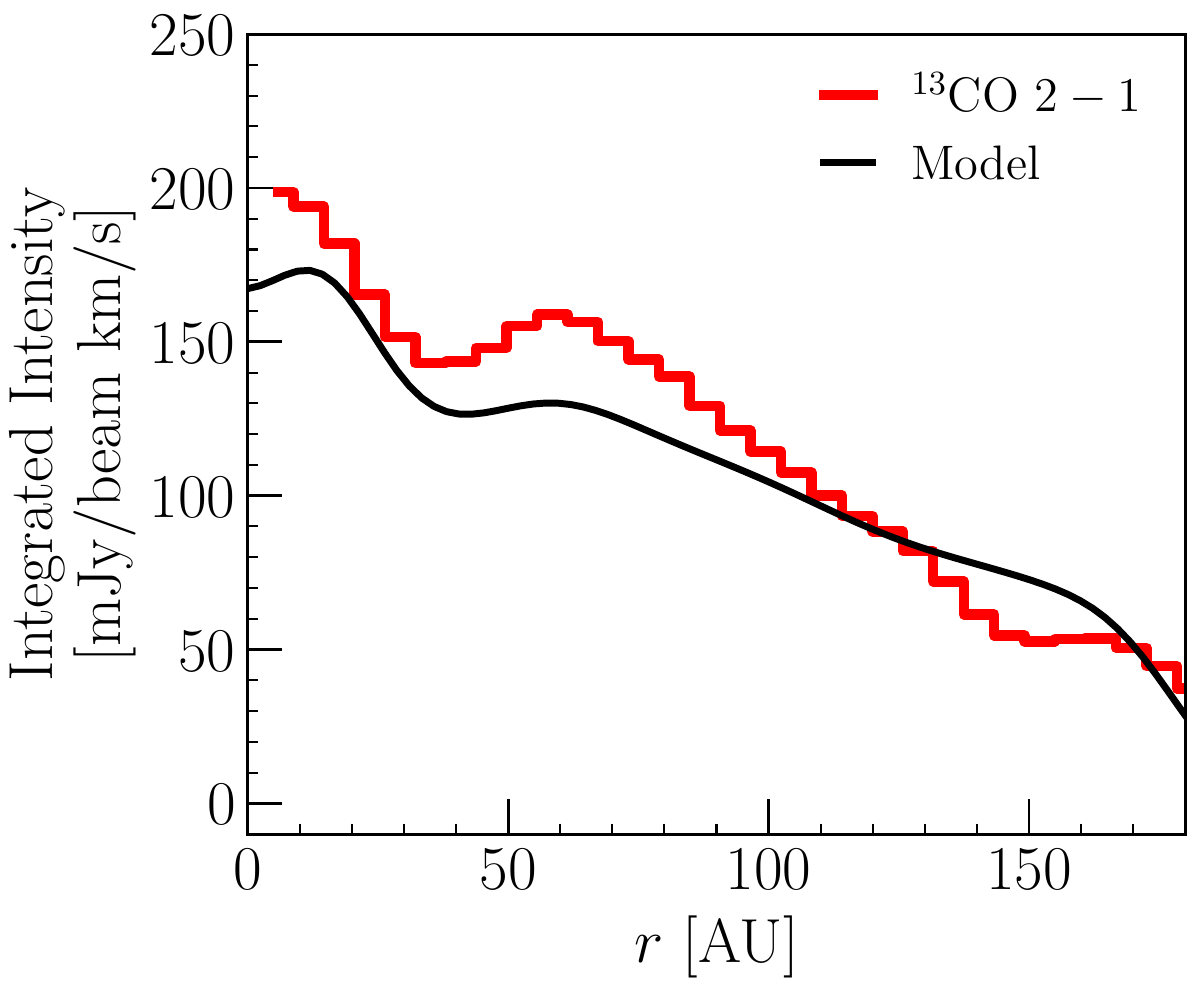}
      \hspace{0.3cm}
      \includegraphics[width=0.3\textwidth,height=0.25\textwidth]{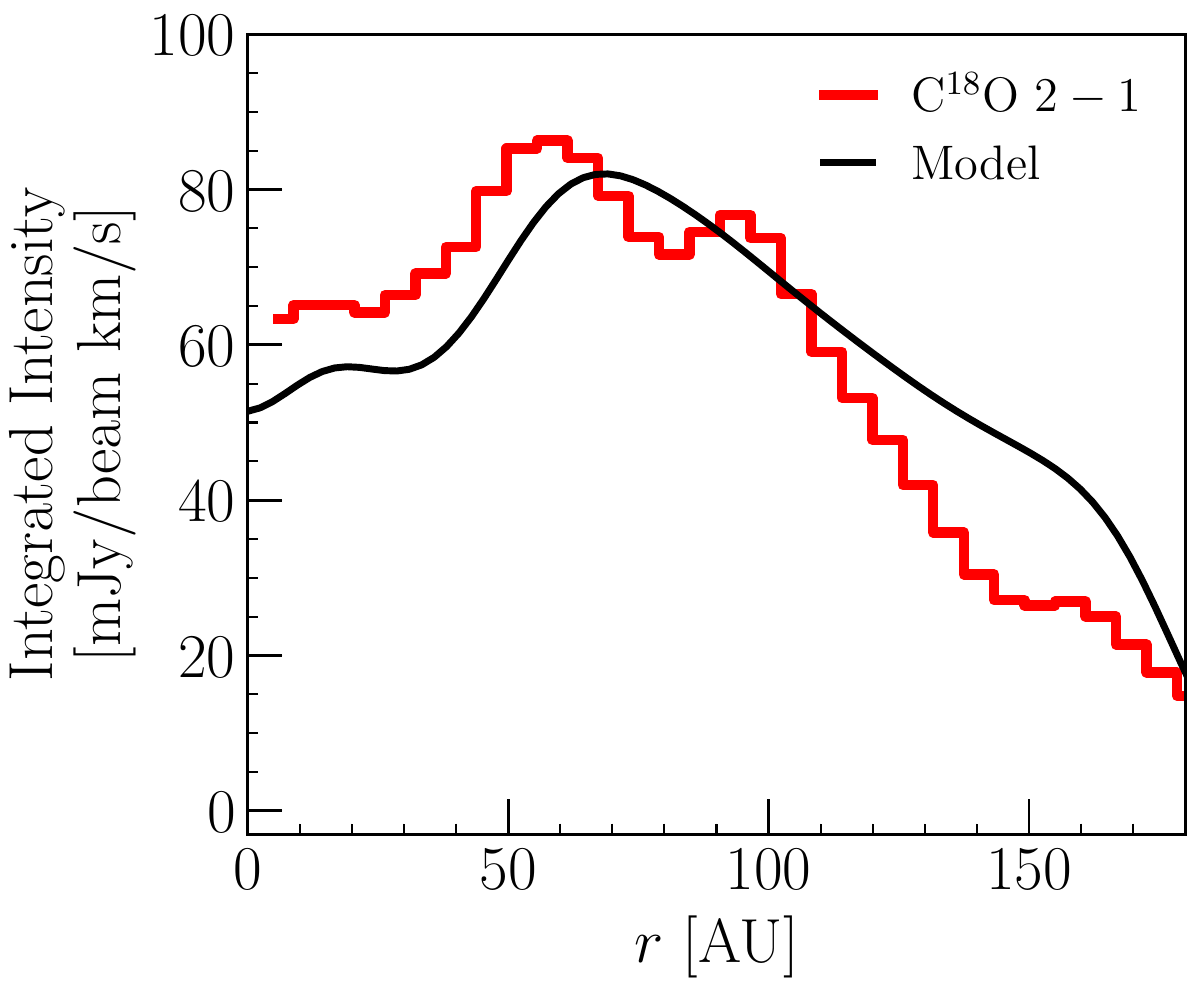}
      \caption{Fits to observables for the optimized \textsc{dali} model. \newchange{Unlike the
      azimuthally-averaged radial profiles in Figure~\ref{fig:radint_dcop_c18o_cont}, the 
      radial profiles shown here are cuts along the disk semi-major axis (PA = 5$^\circ$).}
      ({\it Top left}) Near-infrared to submillimeter spectral energy distribution:
      the model is shown by the black solid curve and the data (red dots) 
      are from \citep{Malfait1998}, IRAS \citep{IRAS1988},
      {\it Herschel} \citep{Pascual2016}, SCUBA \citep{Sandell2011}, and ALMA \citep{Fedele2017a}. 
      ({\it Top middle}) Radial intensity profile of the 1.3 millimeter continuum. 
      ({\it Top right}) Radial intensity profile of $^{12}$CO $J=2-1$. 
      ({\it Bottom middle}) Radial intensity profile of $^{13}$CO $J=2-1$. 
      ({\it Bottom right}) Radial intensity profile of C$^{18}$O $J=2-1$. 
      }
      \label{fig:newmod_fits}
    \end{center}
  \end{figure*}

  \noindent Figure~\ref{fig:mod_dust_struc} shows the change in the dust structure 
  between the fiducial model and our optimized model.
%   \noindent where $R_{\rm dust \ in}$ = 20 AU, the depletion factor $\delta_{\rm dust}$ = 0.27,
%   $R_{\rm gap \ in}$ = 35 AU, $R_{\rm gap \ out}$ = 56 AU, and $R_{\rm dust \ out}$ = 83 AU, and
%   $n_{\rm dust,small}$ describes only the small grain dust population (0.005 -- 1 $\mu$m).
%   \newchange{The ring $R_{\rm dust \ in} - R_{\rm gap \ in}$ (20 -- 35 AU) corresponds to R1 
%   and the ring $R_{\rm gap \ out} - R_{\rm dust \ out}$ (56 -- 83 AU) corresponds to R2 in 
%   Figures~\ref{fig:spec_mom_maps}, \ref{fig:radint_dcop_c18o_cont}, and \ref{fig:dcop_radabun}.}
  The gas density structure of the optimized model remains unchanged
%   and has an inner radius, a depleted region within the dust cavity,
%   a depleted region across the first dust ring and gap,
%   a full region across the second dust ring and outer disk,
%   and an outer radius
% 

  \begin{equation}
  \label{eq:fid_surfdens_gas}
    n_{\rm gas}=\begin{cases}
      0 & \ {\rm for} \  r < R_{\rm gas \ in}\\
      \delta_{\rm gas} \times n_{\rm gas \ cavity} &  \ {\rm for} \ R_{\rm gas \ in} < r < R_{\rm dust \ in} \\
      \delta_{\rm gas} \times n_{\rm gas \ gap}  &  \ {\rm for} \ R_{\rm dust \ in} < r < R_{\rm gap \ out} \\
      n_{\rm gas} & \ {\rm for} \ R_{\rm gap \ out} < r < R_{\rm gas \ out} \\
      0 & \ {\rm for} \ r > R_{\rm gas \ out}. \\
    \end{cases}
  \end{equation}
% 
%   \noindent \change{where $R_{\rm gas \ in}$ = 13 AU, the depletion factor $\delta_{\rm gas,cavity}$ = 0.025,
%   the depletion factor $\delta_{\rm gas,gap}$ = 0.025, and $R_{\rm gas \ out}$ = 180 AU.}

  Adjustments to the dust distribution will affect
  the SED, and the accompanying opacity and temperature variations will affect
  the molecular abundances and thus radial profiles of the $^{12}$CO, $^{13}$CO, C$^{18}$O.
  The new parameters for the small grain (0.005 -- 1 $\mu$m) dust population were chosen
  such that the fit to the SED was maintained and the fits to the CO isotopologues
  were consistent to within a factor of about 10\%. 
    
  The PAH structure was also modified. Originally, the fiducial model dust temperature was set by 
  PAH thermal emission in regions with no dust grains,
  with a global PAH abundance set equal to the ISM abundance.
  PAH abundances have been observed to be low in disks with respect to the ISM
  \citep{Li2003,Geers2006,Thi2014}. However, to achieve the required opacity 
  for a reasonble fit to the $^{12}$CO radial profile for $r$ < 25 AU, it 
  was necessary to keep a high PAH abundance in the inner regions. 
  The PAH abundance in our optimized model was therefore set equal 
  to the ISM abundance for $r \leq 83$ AU and set to 1\% with respect to the ISM abundance for $r > 83$ AU, 
  which is comparable to current estimates for the HD 169142 disk \citep{Seok2016}.
  
  Figure~\ref{fig:newmod_fits} shows the fit to the SED and radial 
  profile of the CO isotopologues from the optimized \textsc{dali} model structure.
  Each point in the SED is well-fit by the \textsc{dali} model with the small grain 
  population modifications. The PAH feature is slightly overproduced because of the high abundance
  in the model at $r \leq 83$ AU. \newchange{Fits to the 
  $^{12}$CO, $^{13}$CO, and C$^{18}$O radial intensity profiles remain consistent 
  with the previous fits from \citet{Fedele2017a} to within 10\%.} \\

%   \newpage
  
  \section{Channel maps}
  \label{app:B}
  
\begin{figure*}[!ht]
    \centering
      \includegraphics[width=\textwidth]{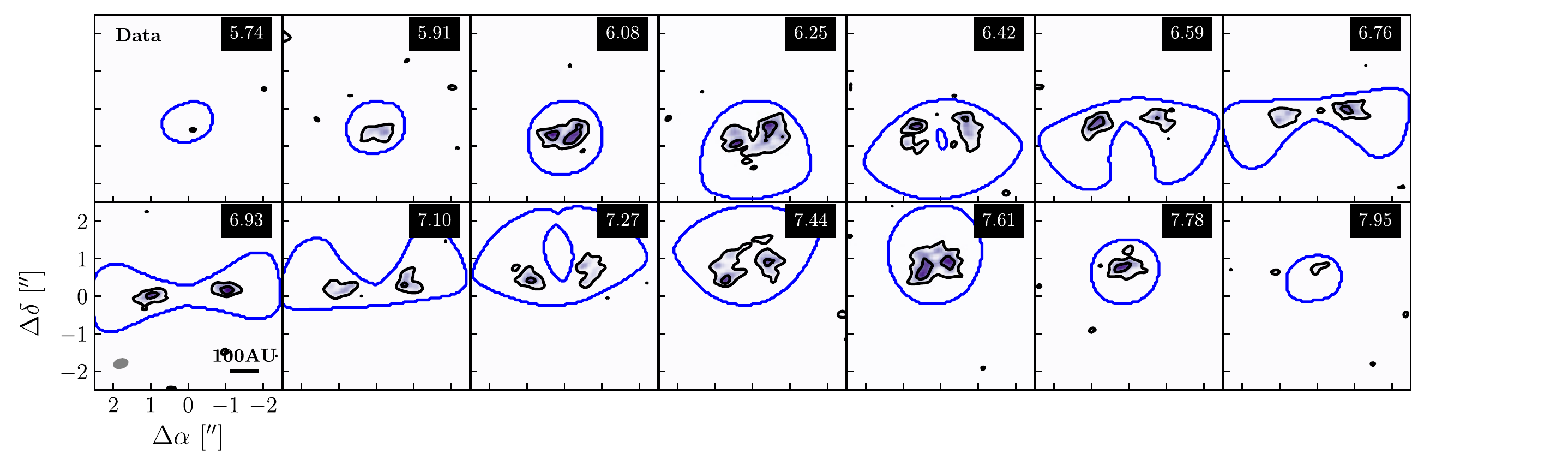}
      \captionof{figure}{DCO$^+$ $J=3-2$ channel maps  from 5.74--7.95 km s$^{-1}$,
      Hanning smoothed to 0.17 km s$^{-1}$ channels. Black contours
      show 5.5 mJy beam$^{-1}$ (1$\sigma$) $\times$ [3, 6, 9]. 
      The blue contours show the outline of the Keplerian mask.
      Channel velocity is shown in the upper-right corner.
      Synthesized beam and AU scale are shown in the lower-left panel.}
      \label{fig:chanmaps}
  \end{figure*}

\begin{figure*}[!htbp]
    \centering
      \includegraphics[width=\textwidth]{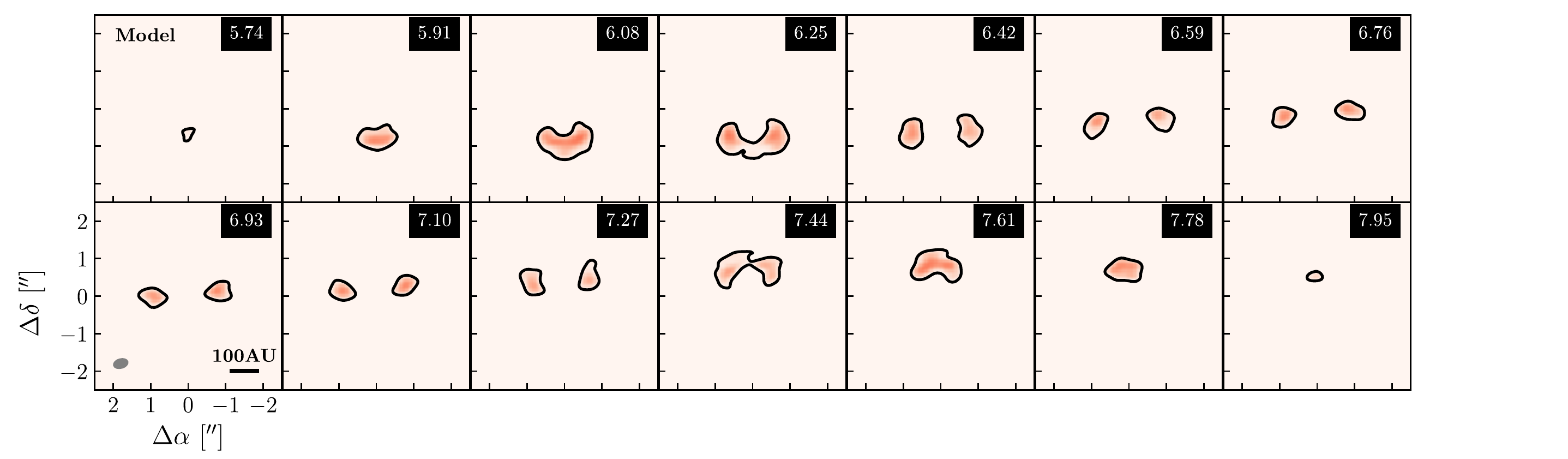}
      \captionof{figure}{DCO$^+$ $J=3-2$ best-fit model channel maps. 
      Black contours show 5.5 mJy beam$^{-1}$ (1$\sigma$) $\times$ [3]. 
      Channel velocity is shown in the upper-right corner.
      Synthesized beam and AU scale are shown in the lower-left panel.}
      \label{fig:chanmaps_model}
  \end{figure*}

\begin{figure*}[!htbp]
    \centering
      \includegraphics[width=\textwidth]{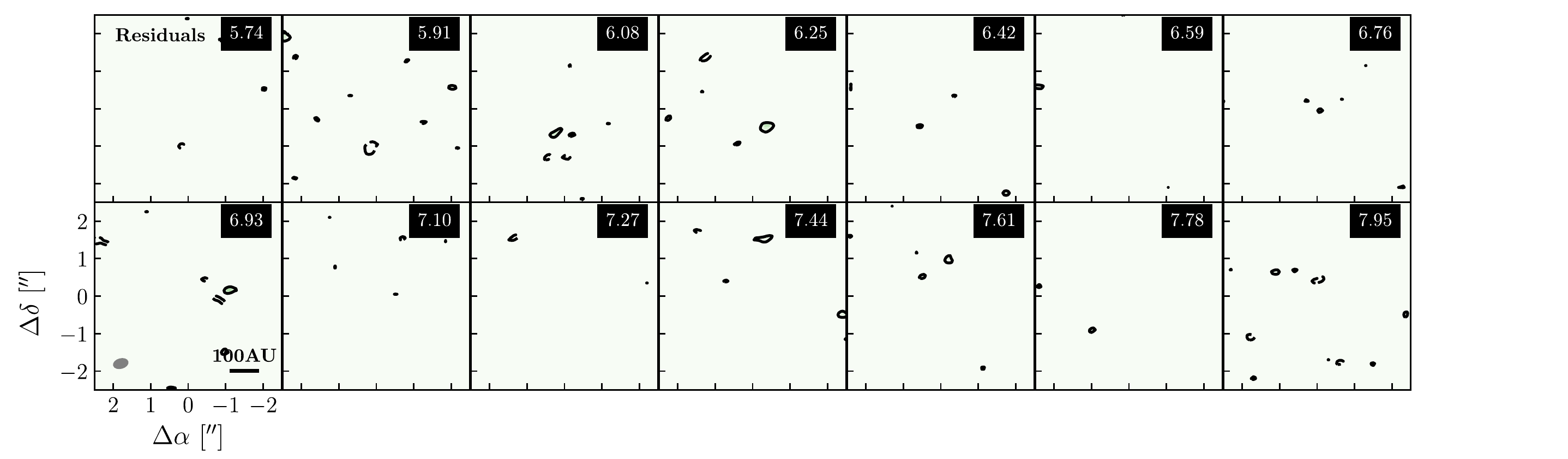}
      \captionof{figure}{DCO$^+$ $J=3-2$ residual (data - model) channel maps. 
      Black contours show 5.5 mJy beam$^{-1}$ (1$\sigma$) $\times$ [3]. 
      Dashed contours are negative at the same intervals.
      Channel velocity is shown in the upper-right corner.
      Synthesized beam and AU scale are shown in the lower-left panel.}
      \label{fig:chanmaps_resid}
  \end{figure*}

  Figures~\ref{fig:chanmaps}--\ref{fig:chanmaps_resid} show channel maps
  of the the DCO$^+$ $J=3-2$ line observed with ALMA and from the modeling presented
  in this paper. Channel maps in Figure~\ref{fig:chanmaps} are
  from the ALMA observations imaged with \textsc{clean} in \textsc{casa}. 
  Channel maps in Figure~\ref{fig:chanmaps_model} are from the model synthetic 
  image cube created with LIME, sampled in the $uv$ plane
  with the \textsc{python} \texttt{vis\_sample} routine, and imaged with \textsc{clean} in \textsc{casa}.
  Channel maps in Figure~\ref{fig:chanmaps_resid} show the residuals of the 
  model subtracted from the data in each velocity channel.
  
\end{appendix}

\end{document}